\documentclass[aps,physrev,preprint,groupedaddress,nofootinbib]{revtex4-2}
\usepackage{graphicx}
\usepackage{amsmath}
\usepackage{amsfonts}
\usepackage{csquotes}
\usepackage{siunitx}    % for pretty formating of quantities

\newlength{\figcolwidth}   % declare a length
\begin{document}
\title{X-rays Emission: a novel tool to detect Extensive Air Showers}

\author{Rodrigo Alberto Torres Saavedra}
\email[Corresponding Author: ]{rodrigo.torressaavedra@gssi.it}
\author{Caterina Trimarelli}
\author{Roberto Aloisio}
\affiliation{Gran Sasso Science Institute, L'Aquila, AQ 67100, Italy}
\affiliation{INFN-Laboratori Nazionali del Gran Sasso, Assergi, AQ 67100, Italy}

\author{John F. Krizmanic}
\affiliation{NASA Goddard Space Flight Center, Greenbelt, MD 20771, USA}
\author{Johannes Eser}
\affiliation{Columbia University, New York, NY 10027, USA}
\author{Austin Cummings}
\affiliation{Pennsylvania State University, University Park, PA 16802, USA}

% \date{\today}

\begin{abstract}
We investigate the feasibility of detecting extensive air showers via their geo-synchrotron X-ray emission from high-altitude platforms. Starting from first principles, we derive a differential expression for the number of emitted photons per unit grammage and photon energy for an ensemble of gyrating shower electrons. The calculation uses noted parameterizations of the electron state variable distributions in the shower to establish a scale for the photon footprint and, further, takes into account the propagation of emitted photons in the atmosphere. The computed fluxes at the position of the detector are used to estimate the detector acceptance and event rate using a bootstrap Monte-Carlo procedure. For a \qty{1}{m}-radius, \qty{70}{\degree} half-aperture circular detector at an altitude between \qtyrange{20}{30}{km} viewing the Earth's limb, we find acceptances at the $\mathcal{O}(\qty{1}{m^2 sr})$ level and integral event rates of $\mathcal{O}(\num{10})$ per month. These results indicate that X-ray geo-synchrotron emission is a promising, complimentary channel for high-altitude indirect cosmic ray detection in the \unit{PeV} regime.

\end{abstract}

\maketitle
\clearpage
\section{Introduction}\label{sec:intro}

The detection of cosmic rays (CR) above \qty{1}{PeV} energies is of chief importance in the efforts to resolve some of the largest long-standing questions in Astroparticle physics~\cite{aloisio_multiple_messengers}. Some of these questions include: elucidating the mechanisms behind the most energetic astrophysical phenomena, identifying the sources and acceleration processes of CR, and characterizing the transition from galactic to extragalactic origins~\cite{snowmass_he_and_uhe_neutrinos,acceleration_and_propagation_of_uhecrs,snowmass_companion_uhecr,cr_selected_topics}, as well as exploring potential secondary emissions produced by Dark Matter decay~\cite{Guepin:2021ljb,Aloisio:2025nts,Aloisio:2007bh}.

Above $\sim \qty{1}{PeV}$, CR detection relies exclusively on ground-based indirect techniques that reconstruct primary CR properties (energy, mass and arrival direction) from the Extensive Air Showers (EAS) they induce in the Earth's atmosphere. Some operating on-ground observatories employ arrays of surface detectors (SDs) such as scintillation counters or water/ice Cherenkov tanks for the direct detection of the secondary particles that reach the ground, as in the case of IceTop~\cite{icetop_all_particle} and GRAPES~\cite{grapes_all_particle}. Other experiments utilize UV telescopes to detect the in-air emission of (primarily) Cherenkov light from the shower secondaries, as in the case of TALE~\cite{tale_all_particle}. Experiments using hybrid detection techniques, such as LHAASO~\cite{lhaaso_all_particle}, combine both approaches to better reconstruct the primary CR properties. At energies larger than $\sim \qty{100}{PeV}$, EAS detection is also possible through their fluorescence emission in the atmosphere \cite{aloisio_multiple_messengers}. Auger~\cite{auger_all_particle, auger_highlights} and TA~\cite{ta_all_particle}, the two largest observatories dedicated to Ultra High Energy Cosmic Rays (UHECR) with energies larger than $\sim$\qty{100}{PeV}, in fact, employ the detection of EAS's fluorescence emission alongside on-ground particles detection.

Current CR data in the energy range between \qty{1}{PeV} and \qty{E3}{PeV} show a general trend on the all-particle energy spectrum of a change in spectral index that defines the CR knee. Nevertheless, estimates of the CR composition in this energy range remain controversial, as they are hindered by model-dependencies in the EAS reconstruction, large systematic errors, and intrinsic uncertainties of the hadronic interaction models~\cite{cr_selected_topics}. Indeed, estimates of the proton flux in the \unit{PeV} range obtained by the KASCADE-Grande Collaboration~\cite{Apel_2013uni} and by the IceTop/IceCube Collaboration~\cite{IceCube_2019hmk} through unfolding of the all-particle CR flux differ by a factor larger than three (as reported in~\cite{shape_of_cr_spectrum}). At higher energies, above \unit{EeV}, Auger and TA data show a larger level of coherence on both spectrum and mass composition \cite{PierreAuger:2025xxv,PierreAuger:2023yym}, underscoring the importance of the hybrid observation technique that combines for the same EAS the detection of on-ground particles and fluorescence emission.

To improve CR observations above PeV energies, in recent years, the detection of EAS from high-altitude platforms (at around 35 km) has attracted growing interest in the astroparticle physics community. With respect to on-ground setups, high-altitude detection offers significant advantages, including broader geometric acceptance, reduced atmospheric absorption, and enhanced detection efficiency. This observational approach has motivated a number of balloon-borne experiments, such as EUSO-SPB2~\cite{euso_spb2}, PUEO~\cite{pueo}, and PBR~\cite{pbr}. At high altitudes, the EAS is observed only through its radiative emissions due to: fluorescence (UV), Cherenkov (visible-UV) and geo-synchrotron processes, the latter provides signals spanning from radio waves up to X-rays (hereafter we will collectively define X-rays as the emission in the energy range above \qty{1}{keV})  \cite{Colin:2006nj,cummings2021modeling1,cummings2021modeling2,raspass_theory,raspass_simulation}. Fluorescence emission, due to the excitation/de-excitation of atmospheric Nitrogen \cite{Colin:2006nj}, has an isotropic nature and can be observed only at the highest energies ($>$ \qty{E2}{PeV}), while Cherenkov and geo-synchrotron emissions, directly produced by the relativistic EAS's secondary particles, are highly anisotropic and can be observed starting from PeV energies \cite{cummings2021modeling1,cummings2021modeling2,raspass_theory,raspass_simulation}.

In the present paper we will concentrate on the X-ray emission from EAS, an emission mechanism that has not yet been explicitly modeled (although some initial work has been presented in~\cite{krizmanic_initial}) and, as we will discuss, can be observed only through high altitude detectors. X-ray emission from EAS is due to the synchrotron process suffered by relativistic electrons (hereafter both e$^+$ and e$^-$), largely dominating the EAS development, interacting with the geomagnetic field. For an electron of Lorentz factor $\gamma$ and pitch angle $\alpha_\text{p}$ (formed by the particle's momentum and the magnetic field direction, see also Fig. 6.1 of~\cite{rybicki2024radiative}), the emitted synchrotron radiation is concentrated in a band about the critical frequency,
\begin{equation}
    \omega_\text{c} = \frac{3}{2} \gamma^3 \omega_\text{B} \sin{\alpha_\text{p}} = \frac{3}{2} \gamma^2 \omega_\text{L} \sin{\alpha_\text{p}} = \frac{3}{2} \gamma^2 \left( \frac{q_\text{e}B}{m_\text{e} c} \right) \sin{\alpha_\text{p}}
\end{equation}
which depends only on the magnetic field strength, $B=\| \vec{B} \|$, electron energy and pitch angle. For typical strengths of the geomagnetic field $\sim \qty{0.5}{G}$, electrons with energies of tens of \unit{MeV} emit in the radio band, $\unit{GeV}$-electrons in the optical, and high-energy (HE) electrons with $E \gtrsim \qty{100}{GeV}$ in the X-ray band. Hence, the X-ray emission is of particular interest as it probes the early stages of the EAS development, when secondary electrons are on average more energetic.

The purpose of this work is to estimate the X-ray signal from EAS that could possibly be observed by a detector at high altitude. In Sec.~\ref{sec:computational_approach} we will derive a differential expression for the number of X-ray photons emitted per unit length (grammage) and per unit of photon energy, following the approach of \cite{cummings2021modeling1,cummings2021modeling2} developed for the EAS's Cherenkov emission. In Sec.~\ref{sec:fluxes}, we will use the EAS's electron distributions to estimate the footprint of the area illuminated on the detection plane and the expected fluxes at the detector. Finally, in Sec.~\ref{sec:acceptance}, we will use this information to provide an estimate of the acceptance and event rate of a simplified detector at high altitude. Conclusions will be then presented in Sec.~\ref{sec:conclusions}.

\section{Computational Approach}\label{sec:computational_approach}

On very general grounds, the energy radiated per unit frequency and solid angle by an accelerated charged particle is proportional to the square of the Fourier transform of its electric field, $\tilde{E}(\omega)$ \cite{rybicki2024radiative}. Hence, to determine the synchrotron emission spectrum one should start by considering the Fourier transform of the retarded fields generated, at the observer position, by a particle gyrating under the effect of an external magnetic field. When the observer is distant, so that the contribution of the radiation field dominates over that of the velocity field, the energy radiated per unit frequency and solid angle by a single particle can be written (in Gaussian units) as \cite{rybicki2024radiative},
\begin{equation}\label{eq:double_differential_energy_distribution_from_first_principles}
 \left ( \frac{dW}{d\omega d\Omega} \right )^{(1)} = c \left|R\tilde{E}(w) \right|^2 = \frac{q^2 \omega^2}{4 \pi^2 c} \left| \int_{-\infty}^{\infty} \hat{n} \times (\hat{n} \times \vec{\beta} ) \exp \left[ i\omega \left(t'-\frac{\hat{n} \cdot \vec{r}_q(t')}{c} \right) \right] dt' \right|^2
\end{equation}
where $\omega$ is the angular frequency of the emitted radiation,  $\hat{n}$ the line-of-sight (LoS) unit vector to the observer at a distance $R$ from the particle, $q$ the particle's charge, and $\vec{\beta}$, $\vec{r}_q$ the particle's velocity and position vectors both evaluated at a given retarded time $t'$.

In literature, the common choice of coordinate system used to compute the radiated energy, as given in Eq.~\ref{eq:double_differential_energy_distribution_from_first_principles}, is to place the coordinate origin at the position of the particle when its retarded time is $t' = 0$. At this instant, by construction, the velocity vector of the particle is along the x-axis, the orbital plane is constrained to the x-y plane, and both the LoS vector and the magnetic field vector are constrained to the x-z plane. Further, two unit polarization vectors are defined: $\hat{\epsilon}_\perp$, along the y-axis in the orbital plane and $\hat{\epsilon}_\parallel$, in the x-z plane, which are two vectors perpendicular and parallel to the magnetic field respectively when projected onto the plane of the observer, i.e., onto a plane normal to the LoS vector. With this convention (refer to Fig. 6.4 of~\cite{rybicki2024radiative}), the vector products of Eq.~\ref{eq:double_differential_energy_distribution_from_first_principles} may be expanded, and the spectrum rewritten as ~\cite{rybicki2024radiative,westfold_synchrotron_polarization},
\begin{equation}\label{eq:double_differential_spectrum_single_particle}
\begin{split}
  \left (  \frac{d W_\perp}{d\omega d\Omega} \right )^{(1)} &= \frac{q^2 \omega^2}{3 \pi^2 c} \left| \frac{a \theta_\mathrm{\gamma}^2}{\gamma^2 c} K_{2/3}(\eta)\right|^2 \\
    \left (\frac{d W_\parallel}{d\omega d\Omega}\right )^{(1)} &= \frac{q^2 \omega^2 \theta^2}{3 \pi^2 c} \left| \frac{a \theta_\mathrm{\gamma}}{\gamma c} K_{1/3}(\eta)\right|^2 
\end{split}
\end{equation}
where we have split the spectrum into its perpendicular and parallel components respectively, and where $\theta$ is the polar angle of the LoS vector with respect to the particle's velocity vector at $t'=0$, $a = \frac{r_\text{L}}{\sin(\alpha_\text{p})}$ is the curvature radius of the particle, $\theta_\mathrm{\gamma}^2 := 1 + (\theta \gamma)^2$, $\eta := \frac{\omega}{2 \omega_\text{c}} \theta_\mathrm{\gamma}^3$, and $K_\mathrm{\nu}$ are modified Bessel functions of the second kind.

Eq.~\ref{eq:double_differential_spectrum_single_particle} can be generalized to the case of an ensemble of $N$ identical particles simply by summing the contribution of each particle to the total electric field and using this field instead. Following \cite{ap_synchrotron_theory_1}, this procedure can be pursued assigning a phase factor $\alpha_j$ to each particle and summing the contributions under the square modulus, as follows,
\begin{equation}\label{eq:double_differential_spectrum_ensemble}
\begin{split}
    \left ( \frac{d W_\perp}{d\omega d\Omega}\right )^{(N)} &= \frac{q^2 \omega^2}{3 \pi^2 c} \left| \sum_j^N \frac{a_j \theta_{\mathrm{\gamma}_j}^2}{\gamma_j^2 c} K_{2/3}(\eta_j)\exp \left\{ i\frac{\omega}{c}a_j\alpha_j \right\} \right|^2 \\
    \left ( \frac{d W_\parallel}{d\omega d\Omega}\right )^{(N)} &= \frac{q^2 \omega^2 \theta^2}{3 \pi^2 c} \left| \sum_j^N \frac{a_j \theta_{\mathrm{\gamma}_j}}{\gamma_j c} K_{1/3}(\eta_j) \exp \left\{ i\frac{\omega}{c}a_j\alpha_j \right\} \right|^2 
\end{split}
\end{equation}
where the summation is over all particles of the ensemble. We must note, however, that in this derivation the assumption of a distant observer is implicit and $\hat{n}$, consequently, is taken to be constant and equal for all particles in the ensemble \cite{ap_synchrotron_theory_1}. We will come back on this assumption discussing the specific case of electrons in the EAS.

The exponential term in Eqs.~\ref{eq:double_differential_spectrum_ensemble} takes into account the possibility of coherent synchrotron emission. In this coherent regime, the energy emitted by an ensemble is proportional to the square of the number of particles in the ensemble, and the power emitted by a single particle cannot be considered independent of that emitted by another particle in the ensemble~\cite{ap_synchrotron_theory_1}. The emission can be considered incoherent only when the frequency of the emitted photons, $\omega$, is always much larger than the ratio of the typical gyration frequency, $\omega_B$,  and the lowest phase shift among electrons, $\Delta\alpha$, that is, $\omega \gg \frac{\omega_B}{\Delta\alpha} =\frac{a}{c\Delta\alpha}$. Given the typical strength of the geomagnetic field, this condition is always satisfied in the case of X-ray emission from an EAS. Therefore, in what follows we will always consider the incoherent case.

In the incoherent regime, the exponential factors under the square modulus in Eqs.~\ref{eq:double_differential_spectrum_ensemble} reduce then to a Kronecker-delta, and the differential spectra for the whole ensemble may be written simply as the sum of the spectra of its constituents,
\begin{equation}\label{eq:double_differential_spectrum_ensemble_not_coherent}
\begin{split}
    \left ( \frac{d W_\perp}{d\omega d\Omega}\right )^{(N)} &= \sum_j^N\frac{q^2 \omega^2}{3 \pi^2 c} \left(  \frac{a_j \theta_{\mathrm{\gamma}_j}^2}{\gamma_j^2 c} K_\text{2/3}(\eta_j) \right)^2 \\
    \left (\frac{d W_\parallel}{d\omega d\Omega}\right) ^{(N)} &= \sum_j^N \frac{q^2 \omega^2 \theta^2}{3 \pi^2 c} \left(  \frac{a_j \theta_{\mathrm{\gamma}_j}}{\gamma_j c} K_\text{1/3}(\eta_j) \right)^2     
\end{split}
\end{equation}
with $a$, $\theta_\mathrm{\gamma}$, $\eta$, and $K_\mathrm{\nu}$ as defined above.

Starting from Eq.~\ref{eq:double_differential_spectrum_ensemble_not_coherent}, we derive the number of photons emitted per unit time by the secondaries at a given stage of the EAS evolution by first computing the power radiated by the electron ensemble, averaged over a full gyration and over all angles. Using the hypothesis of incoherent emission, we can integrate separately over the solid angle each emitting particle, Eq.~\ref{eq:double_differential_spectrum_ensemble_not_coherent} becomes,
\begin{equation}\label{eq:differential_spectrum_ensemble_not_coherent}
\begin{split}
    \left (\frac{d W_\perp}{d\omega}\right )^{(N)} &= \sum_j^N \int_{\Omega_j} \frac{q^2 \omega^2}{3 \pi^2 c} \left(  \frac{a_j \theta_{\mathrm{\gamma}_j}^2}{\gamma_j^2 c} K_\text{2/3}(\eta_j) \right)^2 d\Omega_j\\
    \left (\frac{d W_\parallel}{d\omega} \right) ^{(N)} &= \sum_j^N \int_{\Omega_j} \frac{q^2 \omega^2 \theta_j^2}{3 \pi^2 c} \left(  \frac{a_j \theta_{\mathrm{\gamma}_j}}{\gamma_j c} K_\text{1/3}(\eta_j) \right)^2 d\Omega_j
\end{split}
\end{equation}
where $d\Omega_j = 2 \pi \sin\left(\alpha_{\text{p}, j}\right)\,d\theta_j$ is the strip of solid angle that each gyrating electron illuminates on the sky over one full period of its gyration motion~\cite{rybicki2024radiative}. The total energy then can be obtained by summing the polarization components and carrying out the integral over solid angle, as shown in~\cite{westfold_synchrotron_polarization}, then the expressions in Eq.~\ref{eq:differential_spectrum_ensemble_not_coherent} reduce to,
\begin{equation}\label{eq:total_energy_spectrum_ensemble}
  \left ( \frac{dW}{d\omega} \right )^{(N)} = \frac{\sqrt{3} q^2}{c} \sum_i^N \gamma_i \sin(\alpha_{\text{p},i}) F(x_i)
\end{equation}
where we have introduced the synchrotron function, $F(x) = x \int_x^\infty K_{5/3} (y)\,dy$, and the variable $x_i = \frac{\omega}{\omega_{\text{c},i}}$, which is the ratio between the emitted photon frequency and the critical synchrotron frequency of an electron with a given Lorentz factor and pitch angle.

The regime of X-ray emission by EAS falls in the synchrotron back-reaction regime, because the characteristic energy-loss timescale of the shower electrons is shorter than, or comparable to, their gyration period in the geomagnetic field. As discussed in \cite{ap_synchrotron_theory_1}, the relevant comparison in this regime is between the observational timescale and the electron energy-loss timescale. When the former is much shorter than the latter $-$ as in EAS, where the observation window is typically not larger than hundreds of \unit{ns} $-$ the power emitted by the particle ensemble can be expressed as the energy radiated per gyroperiod, as usually done in the textbook presentations of the synchrotron regime \cite{rybicki2024radiative}. Therefore, using the incoherent hypothesis, we can divide each term in the sum above by the gyration time of the corresponding particle to obtain the total power. Remembering that the gyration period is $T_\text{G} = \frac{2 \pi}{\omega_\text{B}}$ we write then,
\begin{equation}\label{eq:total_power_spectrum_ensemble}
    \left (\frac{dP}{d\epsilon} \right)^{(N)} = \left (\frac{dP}{\hbar d\omega}\right)^{(N)} = \frac{\sqrt{3} q^2}{\hbar c} \sum_i^N \gamma_i \left( \frac{\omega_{\text{B},i}}{2 \pi} \right)\sin(\alpha_{\text{p},i}) F(x_i)
\end{equation}
where $\epsilon$ is the photon energy. Dividing the expression above by the photon energy itself we obtain, finally, an expression for the number of photons emitted per unit time per unit photon energy for a particle ensemble (valid so long as the emission is not coherent and we consider only the emission in a sufficiently small slice of time) as follows,
\begin{equation}\label{eq:total_emitted_number_spectrum_ensemble}
\left (\frac{dN_{\gamma}}{dt\,d\epsilon}\right)^{(N)} = \left (\frac{1}{\epsilon} \frac{dP}{d\epsilon}\right)^{(N)} = \sqrt{3} \alpha_\text{S} \frac{\nu_\text{L}}{\epsilon} \sum_i^N \sin(\alpha_{\text{p},i}) F(x_i)
\end{equation}
where we have simplified the expression by remembering that the gyration frequency is proportional to the Larmor frequency, $\omega_\text{L} = \gamma \omega_\text{B}$, and by introducing the fine-structure constant $\alpha_\text{S}$.

From Eq.~\ref{eq:total_emitted_number_spectrum_ensemble} follows that only two electron state variables play a role: pitch angle and energy, which are entangled in the characteristic synchrotron frequency $\omega_\text{c}$. Therefore, if the ensemble admits a description based on a probability distribution function (PDF) of the population and if the ensemble is large enough, we may write Eq.~\ref{eq:total_emitted_number_spectrum_ensemble} as an integral over the ensemble distribution in terms of these two state variables,
\begin{equation}\label{eq:total_emitted_number_spectrum_continuum_general}
    \left (\frac{dN_\mathrm{\gamma}}{dt\,d\epsilon}\right)^{(N)} = \sqrt{3} \alpha_\text{S} \frac{\nu_\text{L}}{\epsilon} N_\text{e}(t) \int \int dE\,d\alpha_\text{p}\;\frac{dn_\text{e}}{dEd\alpha_\text{p}} \sin(\alpha_\text{p}) F(x)
\end{equation}
where $\frac{dn_\text{e}}{dEd\alpha_\text{p}}$ must be interpreted as the number density of electrons in the ensemble per bin in electron energy and pitch angle, and $N_\text{e}(t)$ is the absolute number of electrons in the shower at the instant $t$. Hereafter, we will assume that the distribution in pitch angle of the electron population may be treated independently of their distribution in energy and, further, that all HE electrons in the shower have approximately the same pitch angle equal to the angle of the shower axis relative to the magnetic field. We will show \textit{a posteriori} (in Sec.~\ref{sec:electron_adf}) that this is an excellent approximation. In this approximation, we can treat the integral over pitch angle as an integral over a Dirac delta distribution obtaining,
\begin{equation}\label{eq:total_emitted_number_spectrum_continuum}
    \left (\frac{dN_\mathrm{\gamma}}{dt\,d\epsilon}\right )^{(N)} = \sqrt{3} \alpha_\text{S} \frac{\nu_\text{L}}{\epsilon} N_\text{e}(t) \sin(\alpha_{\text{p},s}) \int dE\;\frac{dn_\text{e}}{dE}  F(x)
\end{equation}
where $\alpha_{\text{p},s}$ is the pitch angle of the shower-axis relative to the magnetic field. Given the expression above we may write the number of photons emitted by the particles in the shower per unit grammage and photon energy as,
\begin{equation}\label{eq:emitted_photons_master_equation}
    \left (\frac{dN_\gamma}{dX d\epsilon}\right )^{(N)} = \frac{N_\text{e}(X)}{\rho(X) c}\frac{1}{\epsilon} \int_{E_\text{min}} dE \frac{dn_\text{e}(E, X)}{dE} \left (\frac{dP(E, \epsilon, \alpha_{\text{p},s},B)}{d\epsilon}\right )^{(1)}
\end{equation}
where we have used the slant-depth element $d X = \rho dL = \rho c dt$ to express the second derivative in terms of grammage, and $N_\text{e}(X)$ is simply the electrons longitudinal profile of the shower. The lower limit of integration in Eq.~\ref{eq:emitted_photons_master_equation}, selects the relevant electrons' energy range to produce X-ray emission $E_\text{min}=100$~GeV, while for the upper limit we will use a reference value of $1$~PeV, considering that electrons above this energy are too rare to contribute significantly to the emission.

The properties of the emission strongly depend on the characteristics of the electrons distribution in the EAS. Namely, the number of photons emitted per unit grammage and photon energy depend on the electron energy distribution (Eq.~\ref{eq:emitted_photons_master_equation}), whereas the flux of these photons on the extent of the lateral and angular deviations of the electrons in the shower (cf. Sec.~\ref{sec:fluxes}). 
As discussed above, the relevant electron's energy for X-ray emission is in the range $E_\text{e}\gtrsim 100$~GeV, produced mainly during the early stages of the EAS development.

Typically the EAS development is traced by the so-called \enquote{shower-age}, expressed in terms of the slant-depth $X_\text{max}$ corresponding to the maximum of the EAS development (maximum number of electrons) and the generic slant-depth $X$ traversed, $ s=\frac{3X}{X+2X_\text{max}}$. The maximum of the EAS's development is reached at $s=1$ $(X=X_\text{max})$ and the relevant development stages for X-ray emission are in the range $s\lesssim 0.5$. 

Many different authors have studied the distribution of EAS's particles, and nowadays there are very sophisticated Monte Carlo simulations, like \texttt{CORSIKA} \cite{CORSIKA:2025zlj}, whose output can be used to model the electron's distribution in energy, lateral and angular spreads by performing fits of the shower content in small bins of slant depth along the shower development. All approaches to EAS modeling invoke the principle of universality~\cite{lipari_universality} which, briefly, states that the shape of the EAS's particles distributions are independent in form of the primary energy, mass, or shower direction. These quantities, instead, shape the overall characteristics of the distributions, such as normalization, slant-depth position of maximum development and length. As a matter of fact, the universality principle could be scarcely applicable in the early stages of the EAS's development, when the particles content and their phase space distributions depend on the stochasticity of the hadronic interactions. In literature, the applicability of the universality principle to young EAS has not been studied in detail, particularly as the primary interest so far has been in modeling of emission processes occurring near shower maximum where the universality principle definitely holds. We recognize that a dedicated study of the EAS's properties in its early stages is needed but it is beyond the scope of this work. 

\subsection{Longitudinal Profile and Energy Distribution of the Electrons}
\label{sec:longitudinal_profile}

\begin{figure}[htb]
    \begin{minipage}{\figcolwidth}
    \centering
    \includegraphics[width=\linewidth]{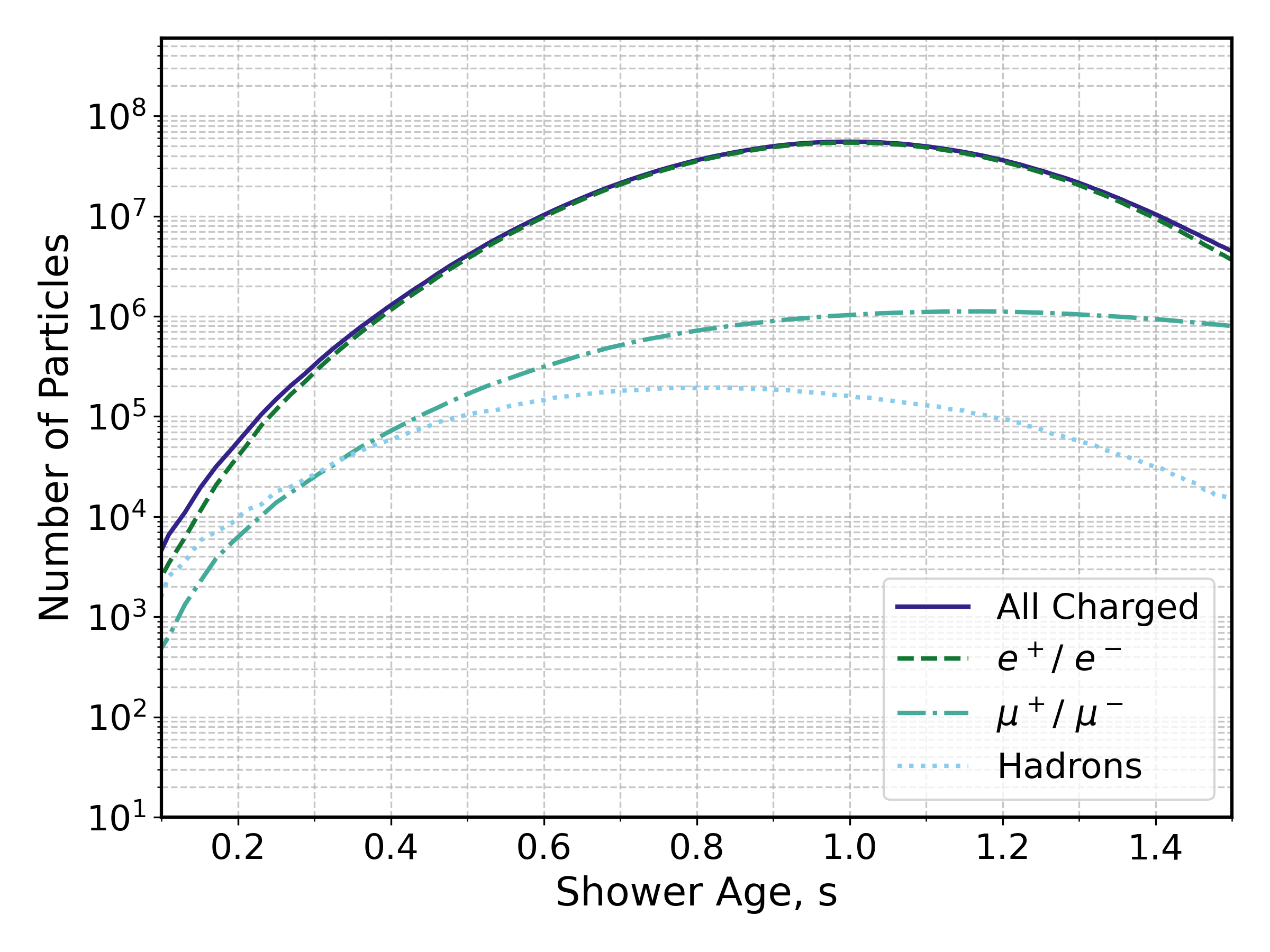}\\
    (a)
    \end{minipage}\hfill
    \begin{minipage}{\figcolwidth}
    \centering
    \includegraphics[width=\linewidth]{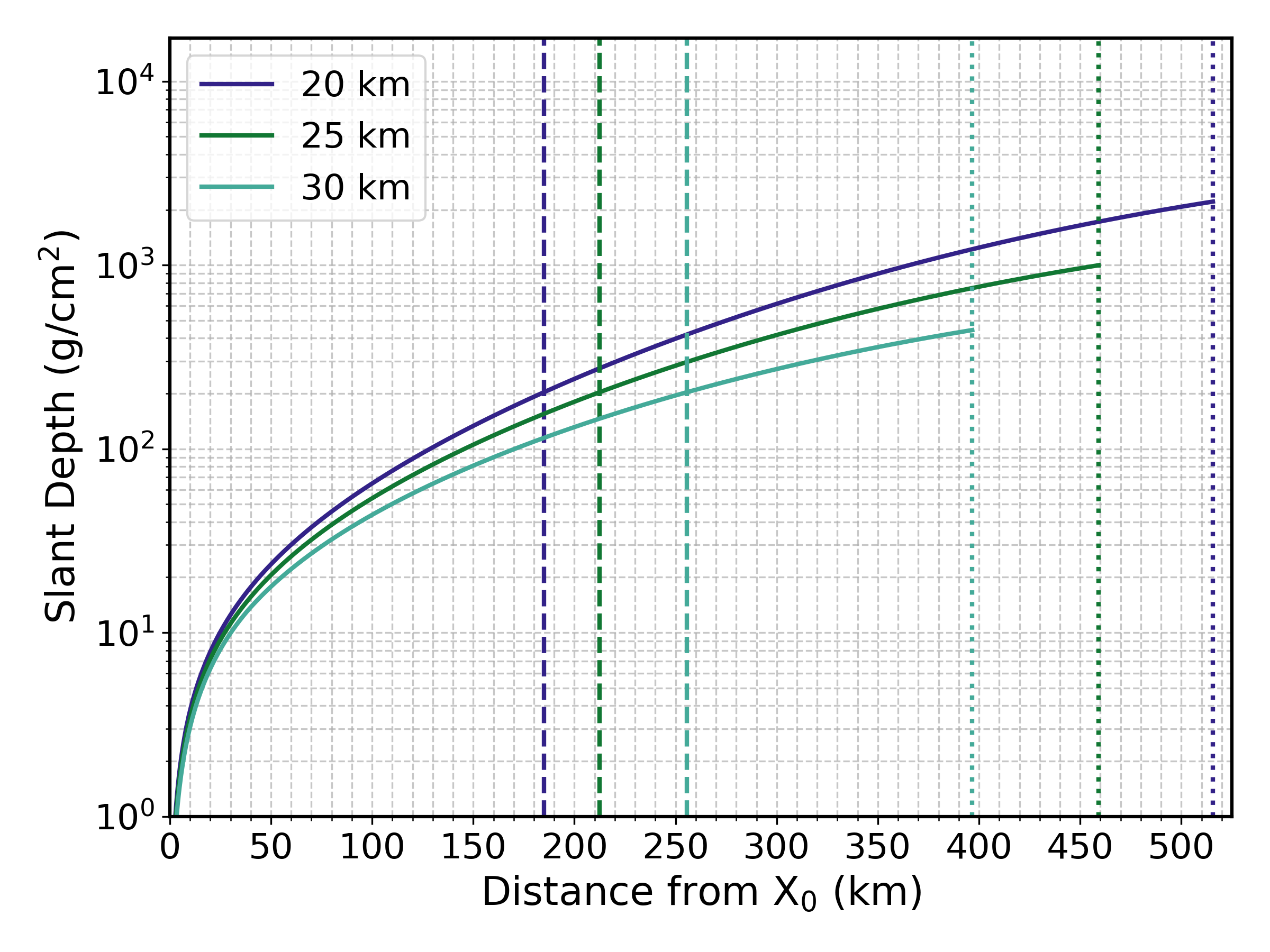}\\
    (b)
    \end{minipage}
    \caption{(a) Longitudinal profile for a \qty{100}{PeV} proton-induced shower as obtained from averaging the output of \num{E3} \texttt{CORSIKA} simulations (see text for details). (b) Examples of the grammage-length relation for a shower developing horizontally in the atmosphere ($\theta_\text{V} = \qty{90}{\degree}$, $X_\text{0} = \qty{30}{gcm^{-2}}$) as observed from three different altitudes. The vertical dashed lines show the distance at which $s = \num{0.4}$ relative to the point of primary interaction, $X_\text{0}$, and the vertical dotted lines show the position of the detector.}
    \label{fig:longitudinal_profile}
\end{figure}

In order to evaluate the electron's longitudinal profile $N_\text{e}(X)$ we use a template profile generated with the \texttt{CORSIKA} simulation framework~\cite{heck1998corsika}, following the approach of~\cite{cummings2021modeling1,cummings2021modeling2}. The template (shown in Fig.~\ref{fig:longitudinal_profile}, left panel) was obtained by averaging a total of \num{1000} simulated proton-induced showers at \qty{100}{PeV}. For different primary energies, the template was rescaled linearly in total electron number and shifted logarithmically in the position of shower maximum using the following parameterization: $\ln{\frac{E}{E_\text{c}}}$ with $E$ the energy of the primary and $E_\text{c} \sim \qty{100}{MeV}$. We note in passing that our simulation framework admits the use of other templates for more accurate modeling of different primary energies and/or masses.

\begin{figure}[htb]
    \begin{minipage}{\figcolwidth}
    \centering
    \includegraphics[width=\linewidth]{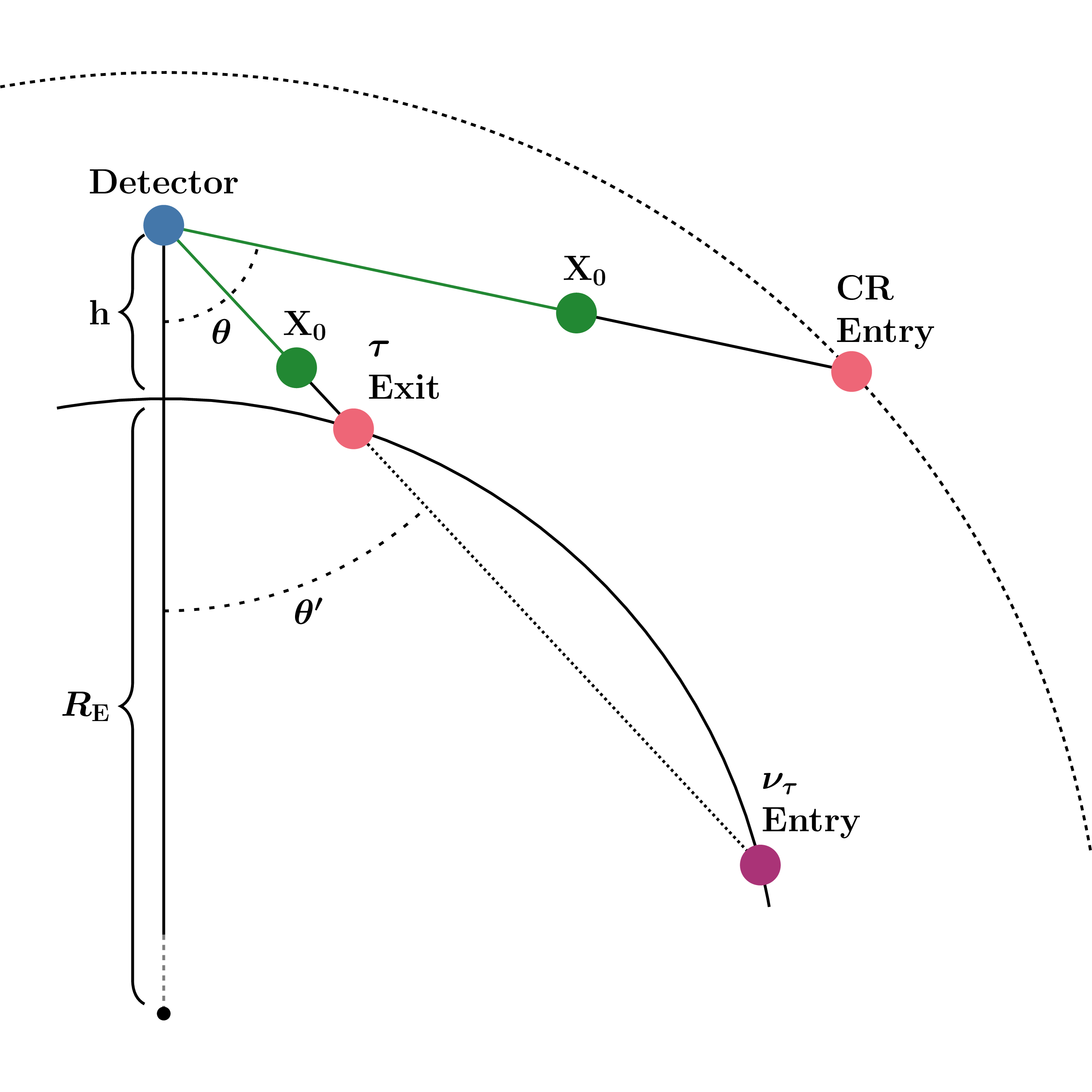}\\
    (a)
    \end{minipage}\hfill
    \begin{minipage}{\figcolwidth}
    \centering
    \includegraphics[width=\linewidth]{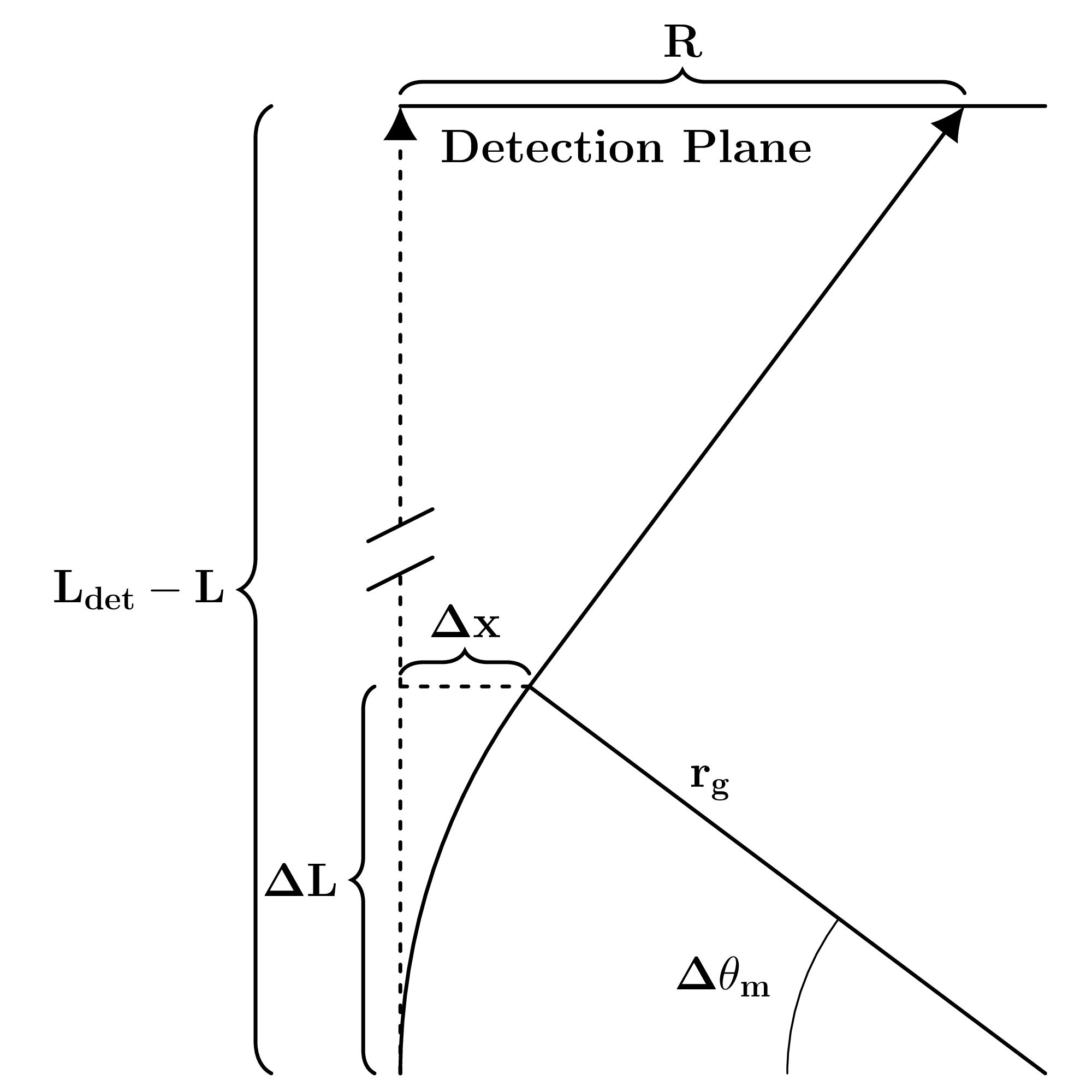}\\
    (b)
    \end{minipage}
    \caption{(a) Geometry conventions used in this work to correlate distances from the point of first interaction (or $\tau$ lepton decay) to altitudes and accumulated grammages. All slant-depths are computed with reference to the CR atmospheric entry or $\tau$ Earth exit points (pink points), offset by the slant-depth at the point of decay or first interaction (green points). Eq.~\ref{eq:master_formula} is solved by integrating numerically along the green portions of the tracks shown in the diagram. The track geometry is entirely defined by the height of the detector, $h$, and the viewing angle, $\theta$ ($\theta'$). See the text and~\cite{cummings2021modeling1,cummings2021modeling2} for further details. (b) Diagram (adapted from~\cite{cummings2021modeling1}) illustrating the effect of considering the electron deviation due to the geomagnetic field and the propagation of the photons along the instantaneous direction of the electron at the point of emission on the extent of the photon footprint area on the detection plane.}
    \label{fig:projection_diagram}
\end{figure}

Slant-depths along the shower axis are assigned using the conventions of \cite{cummings2021modeling1, cummings2021modeling2} (explained also in Fig.~\ref{fig:projection_diagram}). Namely, the shower is developed along the line of sight from the point of atmospheric entry to a detector placed at a fixed altitude, such that the CR trajectory is fully determined by the detector height and the nadir (viewing) angle, $\theta_\text{V}$. Slant-depths along this axis are computed with respect to the first interaction point using the US standard atmosphere density model~\cite{atmosphere1976us}. For EAS induced by Earth-skimming neutrinos~\cite{Alvarez-Muniz_2018owm}, slant-depths are computed instead with respect to the point at which the $\tau$ lepton emerges from the Earth's crust. With these conventions, each point along the shower axis is associated with its altitude, $z$, its distance to the first interaction point, $L$, the distance to the detector, $L_\text{Det}$, and the slant-depth of the first interaction point, $X_\text{0}$. In the right panel of Fig.~\ref{fig:longitudinal_profile} we show a few examples of the grammage-length relation for three trajectories with \qty{90}{\degree} of viewing angle as observed by a detector placed at \qty{20}{km}, \qty{25}{km}, and \qty{30}{km}.

For the electron distribution, the $\frac{dn_e(E, X)}{dE}$ term in Eq.~\ref{eq:emitted_photons_master_equation}, we will adopt a simplified approach by extending the Nerling model \cite{nerling2006universality} to shower ages $s\lesssim 0.5$. This model reportedly provides a better fit to simulations of hadronic showers with UHE primaries simulated with \texttt{CORSIKA} with respect to the models of Hillas~\cite{hillas1982angular} and Giller~\cite{giller2004energy}. Moreover, the Nerling model has been explicitly validated down to $s=0.8$ and up to $E_\text{e} = \qty{10}{GeV}$ \cite{nerling2006universality}, thus it will be extrapolated by \num{0.3} units in shower age (on average) and by two orders of magnitude in electron energy (at most).

The EAS electron energy distribution in the Nerling model is given as a function of the particles kinetic energy $E_\text{k}$ by,
\begin{equation}\label{eq:nerling}
\frac{dn_\text{e}(E_\text{k}, X)}{dE} = \frac{1}{N_\text{e}}\frac{dN_\text{e}(E_\text{k},X)}{dE} \propto \frac{a_\text{0}}{(E_\text{k} + a_\text{1}) (E_\text{k}+a_\text{2})^s}
\end{equation}
where $a_\text{0}, a_\text{1}$ and $a_\text{2}$ are model parameters that depend on the shower age $s$ and are tabulated in~\cite{nerling2006universality}. In the original approach, using the normalization constant provided by Nerling in Eq.~\ref{eq:nerling}, we have observed sizable violations of the unitarity condition for shower ages $\leq$ \num{0.4}. To overcome this difficulty, guided by the fact that at early shower ages electrons in the EAS can get a sizable fraction of the primary energy, we have assumed a normalization condition such as,
\begin{equation}
    \int_{50\text{keV}}^{E_\text{max}} \frac{dn_\text{e}(E_\text{k},X)}{dE} dE_\text{k} = 1
\end{equation}
where $50$~keV is the lowest energy threshold adopted in the \texttt{CORSIKA} simulation, and for the upper bound of integration we take \qty{1}{\%} of the primary particle energy $E_\text{max}=0.01\times E_\text{CR}$.

\subsection{Atmospheric Effects}\label{sec:atmosphere}

\begin{figure}[htb]
    \begin{minipage}{\figcolwidth}
    \centering
    \includegraphics[width=\linewidth]{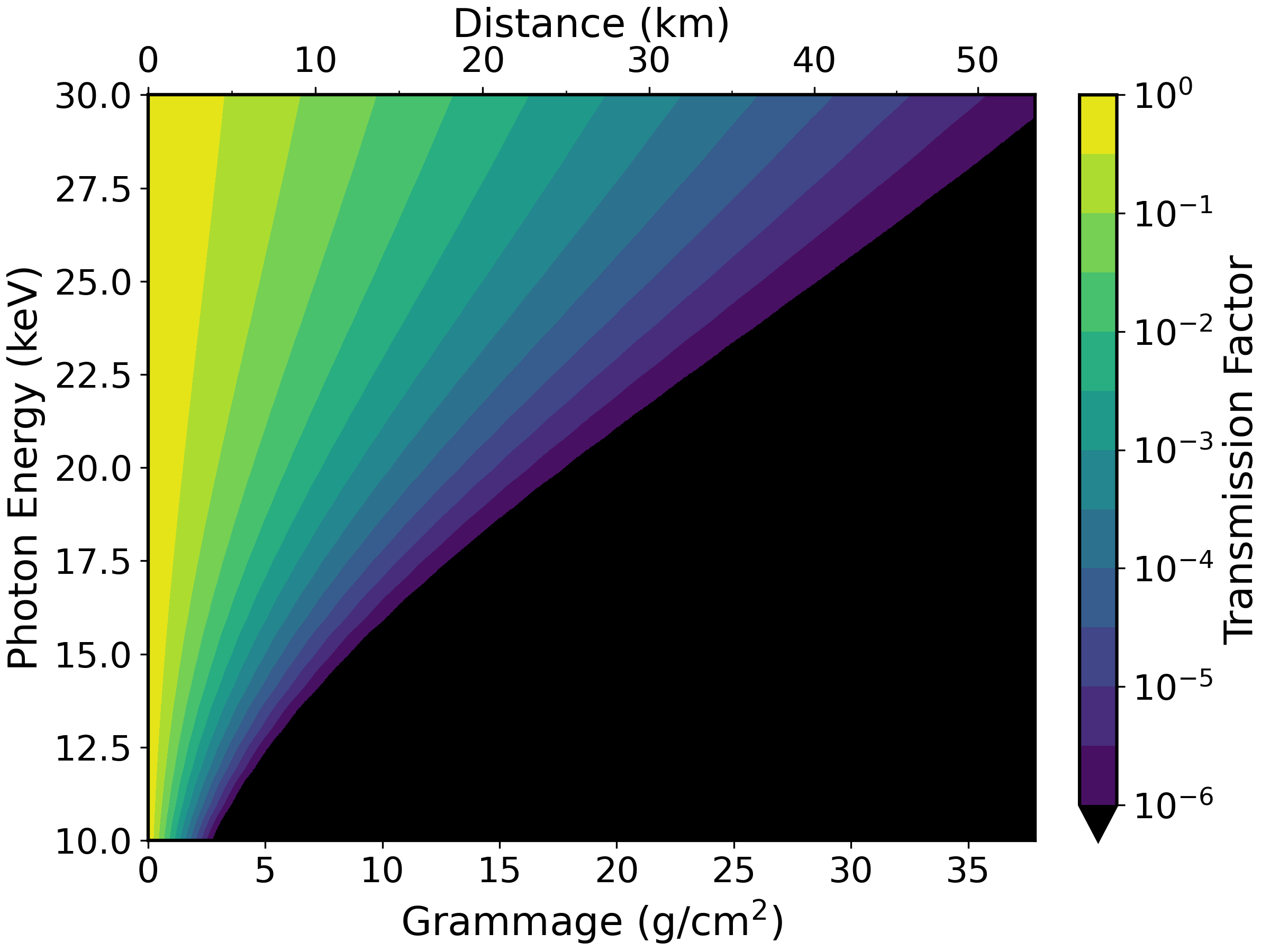}\\
    (a)
    \end{minipage}\hfill
    \begin{minipage}{\figcolwidth}
    \centering
    \includegraphics[width=\linewidth]{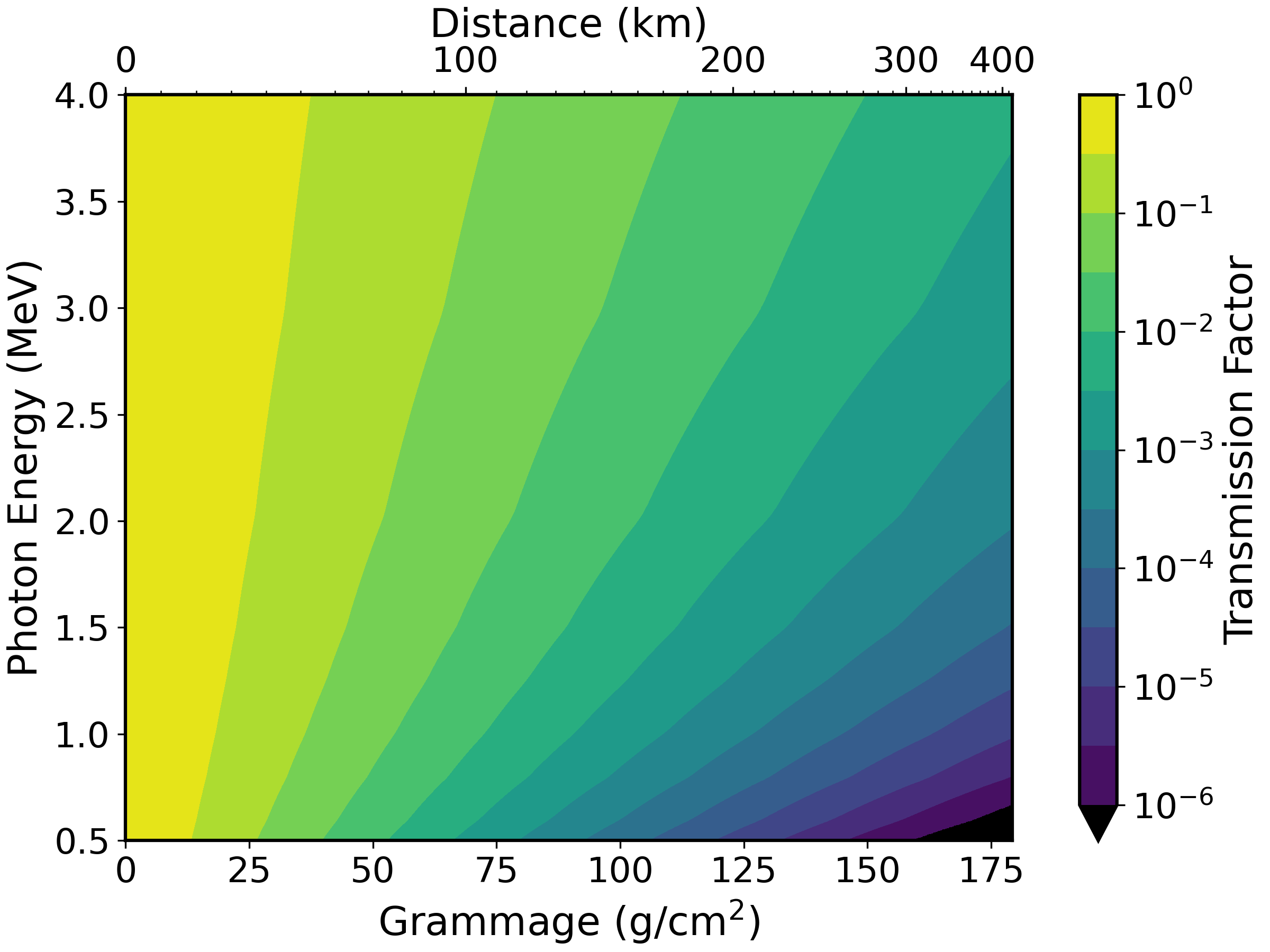}\\
    (b)
    \end{minipage}
    \caption{(a) Contour plot of the transmission factors from the point of emission to the position of a detector at $z = \qty{36}{km}$, computed according to Eq.~\ref{eq:attenuation_factor}, for photons in the \qtyrange{10}{30}{keV} energy band. (b) The same but for photons in the \qtyrange{0.5}{4.0}{MeV} energy band.}
    \label{fig:transmission_factors}
\end{figure}

In general, the number of photons reaching the detector will not coincide with the number of photons emitted during shower development, since photons are subject to absorption and scattering in the atmosphere through several processes such as photoelectric absorption, Compton scattering, and pair production. The extent of attenuation is quantified by the mass attenuation coefficient, $\mu_\text{a}$, which is itself derived from the photon cross section for the different absorption/scattering mechanisms. With these coefficients, the transmission factor between the emission point and the detector (or, alternatively, the survival probability) is calculated as,
\begin{equation}\label{eq:attenuation_factor}
    \text{Tr}\,(E_\mathrm{\gamma}, \Delta X) = \displaystyle \exp \left[ -\int_{L_\text{Det} - L}^{L_\text{Det}} \mu_\text{a} (E_\mathrm{\gamma}) \, \rho(z(l)) dl  \right] = \displaystyle \exp \left[ - \mu_\text{a} (E_\mathrm{\gamma}) \Delta X  \right]
\end{equation}
where $z$, $L$, and $L_\text{Det}$ all follow from the definitions in Sec.~\ref{sec:longitudinal_profile}, and $\Delta X$ is the grammage difference between the point of emission and the detector position.

To evaluate the transmission factors in Eq.~\ref{eq:attenuation_factor}, we obtained the mass attenuation factors from the NIST XCOM database~\cite{nist} for a gas mixture with the same composition as the US standard atmosphere~\cite{atmosphere1976us}, assuming ideal gas behavior such that the volume and mole fractions of its constituents coincide. For illustration, transmission factors were computed for a reference CR trajectory with \qty{90}{\degree} viewing angle and a detector altitude of \qty{36}{km} (an altitude within the typical range for balloon experiments), and are shown in Fig.~\ref{fig:transmission_factors} as a colormap plotted against both the photon energy and $\Delta X$. Figure ~\ref{fig:transmission_factors} shows that even at sub-orbital altitudes, X-rays with energies $\sim \qty{10}{keV}$ are strongly absorbed in the atmosphere, whereas gamma-rays may travel for $\gtrsim \qty{100}{km}$ before being absorbed.

Taking into account the effects of photons propagation through the atmosphere, the total number of photons transmitted to the detection plane is obtained by integrating the photon emission spectra (Eq.~\ref{eq:emitted_photons_master_equation}) corrected by the transmission factor (Eq.~\ref{eq:attenuation_factor}) over the full shower development, as follows,
\begin{equation}\label{eq:master_formula}
\frac{dN_\mathrm{\gamma}}{dE_\mathrm{\gamma}} = \int_{X_\text{min}}^{X_\text{max}} dX \;\text{Tr}\,(E_\mathrm{\gamma}, X_\text{Det} - X) \int_{E_\text{min}}^{E_\text{max}} \frac{dN_\mathrm{\gamma}}{dX dE_\mathrm{\gamma}}     
\end{equation}
where, as discussed above, the integration limits in the electron energy $E$ are taken from \qty{100}{GeV} up to the upper bound of the electron's energy distribution normalization interval, encompassing all electrons that contribute to the synchrotron emission in the \qty{10}{keV} to \qty{10}{MeV} photon energy band. The integration limits in the slant-depth $X$, instead, extend from the first interaction point between the primary CR and the atmosphere, $X_0$, up to the intersection between the EAS axis and the plane of the detector (cf. Sec.~\ref{sec:longitudinal_profile}).

\section{Fluxes}\label{sec:fluxes}

The photon flux at the detection plane is determined by the spatial distribution of the arriving photons, which, in turn, is influenced by several physical phenomena. In particular, at each slant-depth value along the shower development, the emission footprint depends on the electrons' lateral deviations away from the shower core (cf. Sec.~\ref{sec:electron_ldf}). At each instantaneous electron position, the photons are emitted in an extremely narrow cone about the instantaneous electrons' velocity vector, whose half-aperture is proportional to the inverse of the electrons' Lorentz factor. Consequently, the photon directions are strongly dependent on the distribution of the electrons' polar angle away from the shower-axis (cf. Sec.~\ref{sec:electron_adf}). Further, the electrons' lateral and angular distribution are modified by the geomagnetic field. While it is true that HE electrons have Larmor radii $\sim \mathcal{O}(\num{E4})$ \unit{km} in a \qty{0.5}{G} field, horizontal showers that develop at high-altitude in a rarefied atmosphere can develop over distances of over $\sim \mathcal{O}(\num{E3})$ \unit{km}, calling for a quantitative study of this phenomenon (cf. Sec.~\ref{sec:electron_ldf}). Lastly, the emission footprint at the point of emission will generally differ from that at the point of detection due to the effects of photon propagation and projection (cf. Sec.~\ref{sec:areas}). It is the purpose of this section to investigate each of these effects in more detail.

\subsection{Angular distribution}\label{sec:electron_adf}

\begin{figure}[htb]
    \centering
    \includegraphics[width=\dimexpr 1.5\figcolwidth\relax]{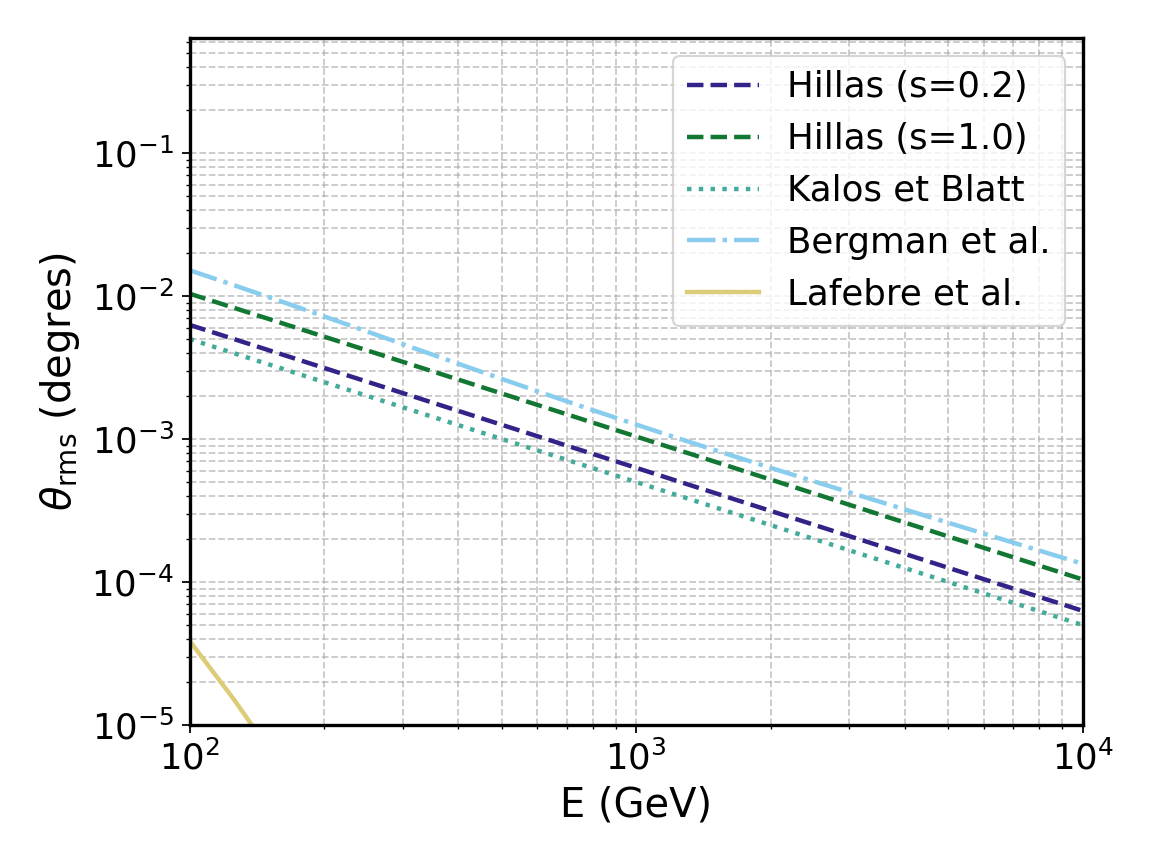}
    \caption{Comparison of the RMS of the electron's polar angle provided by different models (as labeled) of the Angular Distribution function of high energy electrons in a \qty{100}{PeV} EAS.}
    \label{fig:adf_comparison_low_energy}
\end{figure}

As for the electron energy distribution (cf. Sec.~\ref{sec:longitudinal_profile}), an angular distribution function (ADF) is needed that is valid for hadronic showers, HE electrons, and young shower ages. We have compared a total of four different ADF parameterizations: the Hillas parameterization~\cite{hillas1982angular} based on fits to dedicated Monte Carlo simulations of pure electromagnetic showers with energies up to \qty{1}{TeV}; the Kalos and Blatt parameterization~\cite{kalos1954angular} based on an approximation to the solution of the particle cascade integro-differential equation; the Lafebre parameterization~\cite{lafebre2009universality} which is based on fits to simulated electron distributions (in the energy range from \qtyrange{1}{E3}{MeV}) obtained with \texttt{CORSIKA}; the Bergman parameterization~\cite{bergman2013efficient} which also fits simulated showers with \texttt{CORSIKA} but focusing on electrons in the \qtyrange{1}{E3}{GeV} energy range. A comparison of the different models at the relevant electrons' energy ($E\ge 100$~GeV) is shown in Fig.~\ref{fig:adf_comparison_low_energy}. As can be seen from Fig.~\ref{fig:adf_comparison_low_energy}, all models, with the exception of the Lafebre parameterization, agree in their predictions of the electrons' root-mean-square (RMS) angle with respect to the shower-axis to within a factor of $\sim \num{2}$, particularly for HE electrons and in spite of the different validity domains of these models. For consistency with our choice of the electron energy distribution, we could have used the Nerling ADF parameterization~\cite{nerling2006universality}, however, this model is a parameterization for the angular distribution of emitted Cherenkov photons. While it is true that the photon distribution closely follows that of the electrons in the shower, it is convolved with the Cherenkov angle of emission, which is much larger than the typical emission angle of emission for X-ray synchrotron photons.

In the present paper, we adopt the Bergman distribution~\cite{bergman2013efficient}, as it is the only parameterization that explicitly addresses our case of HE electrons in a HE hadronic shower. We highlight too that for HE electrons, the RMS polar angle with respect to the shower axis is $\mathcal{O}(\num{E-3})$ degrees, which provides an \textit{a posteriori} justification of the approximation that HE electrons travel in directions nearly parallel to the shower-axis.

\subsection{Lateral distribution}\label{sec:electron_ldf}

The conventional description of spatial electron densities in codes that simulate EAS secondaries and their electromagnetic emissions rely on fine-tuned Nishimura-Kamata-Greisen (NKG) functions~\cite{kamata1958lateral,greisenlateral}. These distributions are often parameterized in terms of a dimensionless variable, $x = \frac{r}{r_\text{M}}$, defined as the ratio the electrons' lateral distance from the EAS axis, $r$, to the Molière radius, $r_\text{M}$. The latter effectively sets the maximal scale of lateral deviations and depends only on the local atmospheric density. At sea level, $r_\text{M} = \qty{78}{m}$, while at high altitudes altitudes ($>$~\qty{20}{km}), where the atmosphere is rarefied, this scale can reach values on the order of several kilometers. The Molière radius is defined with respect to the population of LE energy electrons in the shower, with energies near the critical energy of air $E_c \sim \qty{100}{MeV}$ \footnote{The critical energy of air is the energy at which the electrons' losses due to bremsstrahlung and ionization become equal.}, close to shower maximum. By definition, it encloses \qty{90}{\percent} of the energy content at shower maximum.

For young showers ($s \sim \num{0.5}$), one would expect a smaller displacement from the EAS axis respect to $r_\text{M}$. This simple intuition is based on the physics of the pair-production and bremsstrahlung processes that, driving particle multiplication in the shower, are boosted forward for HE electrons. At low shower ages electrons have undergone less multiple Coulomb scatterings (MCS) and have therefore a narrower angular spread. Moreover, electrons have propagated over shorter path lengths, consequently, their angular spread is not yet translated into significant lateral displacements. This hypothesis is confirmed by the behavior of the Lafebre parameterization of the lateral distribution function (LDF) within its domain of validity (\qtyrange{1}{E3}{MeV}), which to our knowledge is the only existing multivariate description of the LDF as a function of both electron energy and shower age \cite{lafebre2009universality}. We did not attempt to extrapolate Lafebre's distributions beyond \qty{1}{GeV} in electron energy as they agreed poorly with other distributions in the high-energy regime (see Fig.~\ref{fig:adf_comparison_low_energy}).

For the lateral scale then, we adopt a hybrid approach using Hillas' parameterization of $\langle x \rangle$~\cite{hillas1982angular}, and the RMS polar angle according to the Bergman distribution~\cite{bergman2013efficient}. In particular,
\begin{equation}\label{eq:hillas_lateral_scale}
    \langle x \rangle = (2.05 + 2.56 s^2)(E_\text{k} - 7)^{1/4} w^{1/2} \frac{21}{E_\text{k}}
\end{equation}
where $E_\text{k}$ is the kinetic energy of the electron in MeV, and $w = w(\theta)$ is a scaled angular variable function of the electrons' polar angle $\theta$, defined as,
\begin{equation}\label{eq:hillas_scaled_angular_variable}
    w(\theta) = 2 (1 - \cos\theta)\left(\frac{E_\text{k}}{21}\right)^2 \simeq \left(\frac{\theta E_\text{k}}{21}\right)^2
\end{equation}
taking $\theta=\theta_{\text{RMS}}(E)$ the lateral deviation is a function of the electron's energy through the RMS of the angular deviation, which we take according to the Bergman parameterization.

The electrons later deviation is also affected by the geomagnetic field. To model this effect we follow the Hillas approach~\cite{hillas1982angular}, based on the definition of the quantity $x_\text{m}$ that identifies the effective path integral over which the magnetic field has acted on an electron of energy $E$, given its whole particle history, which was obtained simply from rearranging the Lorentz force equation and tabulating the values observed in the Monte Carlo simulation of this quantity, as defined below,
\begin{equation}
    x_\text{m} = E \int \pm \frac{dx}{U(x)}
\end{equation}
where $U(x)$ is the instantaneous energy of the particle along its entire history during the shower development, and the integral must be interpreted as a summation of several integrals each with a sign assigned depending on the charge of the particle ancestor. With this definition, Hillas provides an analytic fit for the expectation value given by,
\begin{equation}\label{eq:lat_magnetic}
    \langle x_\text{m} \rangle = \frac{30 w^{0.2}}{1 + 36 / E} (gcm^{-2})
\end{equation}
where $w$ is the scaled angular variable introduced in Eq.~\ref{eq:hillas_scaled_angular_variable}. Using the typical quantities for HE electrons ($E > \qty{100}{GeV}$) in the shower we find that the expected path integral is on the order of one radiation length and, consequently, the expected lateral deviations are on the order of a few meters to tens of meters at most.

\subsection{Area Estimates}\label{sec:areas}

Taking into account the effect of the angular and lateral distributions, alongside magnetic field deviations, we can define a procedure to estimate the area illuminated by the synchrotron photons on the detector plane. We first subdivide the length between the point of first interaction and the detector position into bins of equal grammage (see Fig.~\ref{fig:projected_footprint}). A bin width of \qty{0.1}{g cm^{-2}}, was chosen so that the corresponding bin length at the highest altitudes ($\sim \qty{35}{km}$) was on the order of $\qty{100}{m}$. For every bin in grammage, we compute the shower age, the corresponding average electron energy and the expected angular RMS from the Bergman distribution. Then using Eq.~\ref{eq:hillas_lateral_scale} and \ref{eq:lat_magnetic} we compute the total later deviations. The procedure above returns an estimate for the expectation for the angular and lateral deviations of the electrons for a given stage of shower development. These electrons will emit photons with directions nearly tangential to their momentum, and thus, to properly estimate the area illuminated on the detection plane, the direction of the photons needs to be projected from the point of emission to the intersection with the detection plane, as shown in the right panel of Fig.~\ref{fig:projection_diagram}.

After projecting the photon directions, we obtain an estimate of the illuminated area per bin in grammage of shower development, or rather, an array of illuminated area values. To obtain an effective area, a weighted average is performed using the electron energy distribution (re-normalized in the interval of energies contributing to the observation band of interest) as a weighing function. Then we perform a second weighted average of the areas over shower age, using the number of transmitted photons as a weighing factor. The value obtained after this procedure is the area on the detection plane used to compute the expected fluxes discussed in the next section.

\subsection{X-ray fluxes from EAS geo-synchrotron}\label{sec:fluxes_results}

\begin{figure}[htb]
    \begin{minipage}{\figcolwidth}
    \centering
    \includegraphics[width=\linewidth]{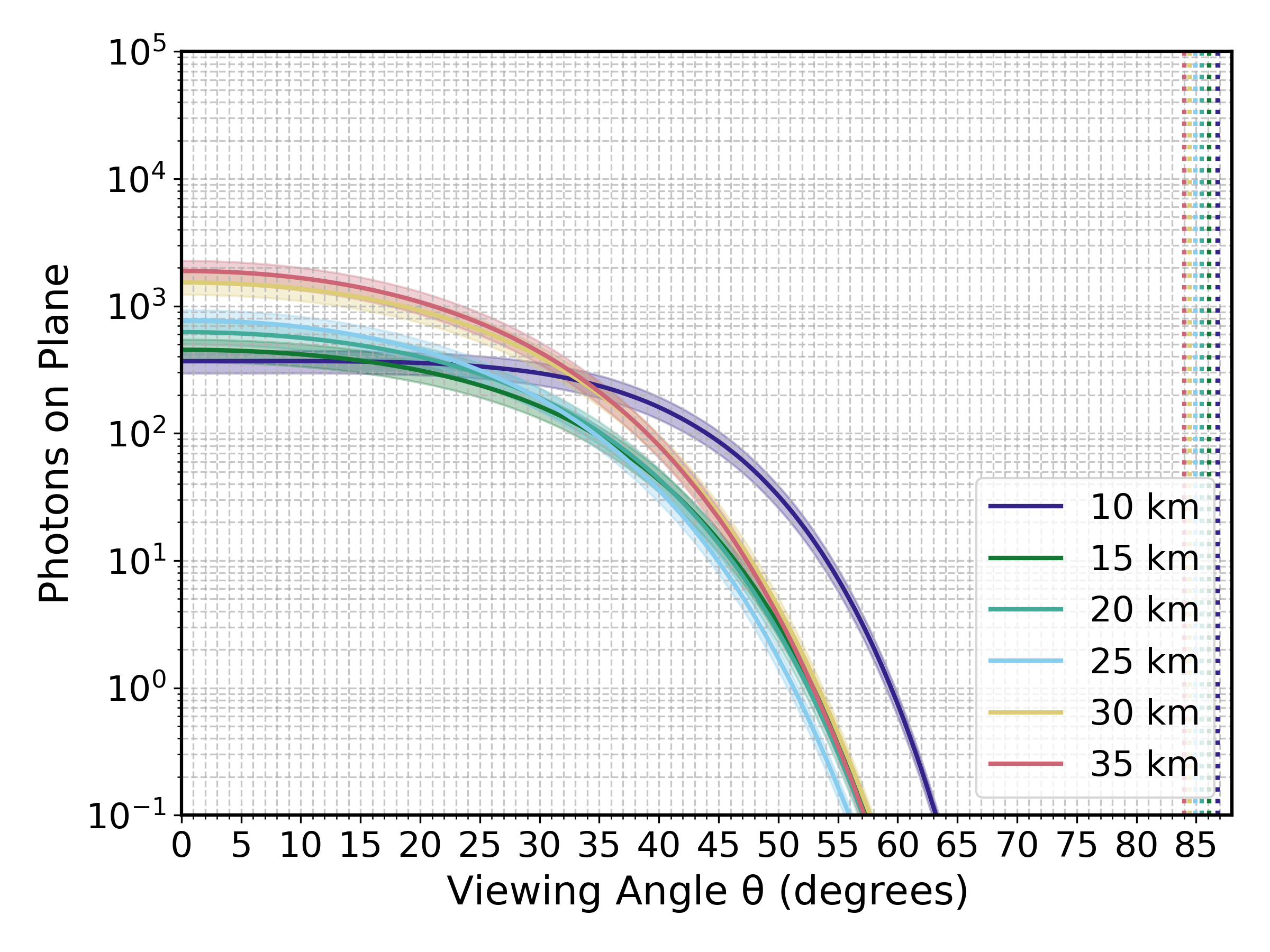}\\
    (a)
    \end{minipage}\hfill
    \begin{minipage}{\figcolwidth}
    \centering
    \includegraphics[width=\linewidth]{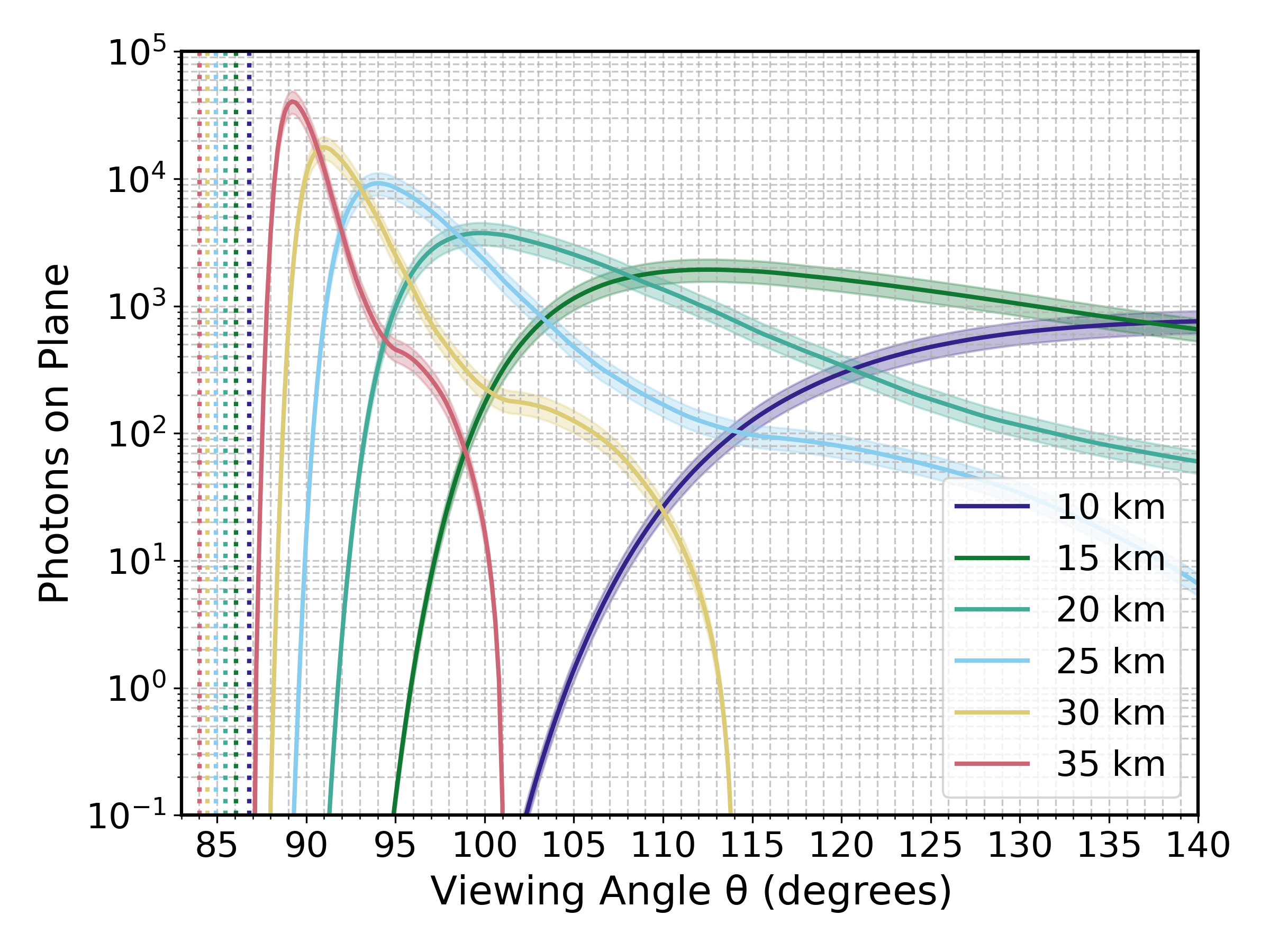}\\
    (b)
    \end{minipage}
    \caption{(a) Number of photons across all photon energy bands (\qtyrange{10}{E4}{keV}) arriving at the position of a detector at several altitudes for below-the-limb events from \qty{100}{PeV} proton-induced EAS. (b) The same, but for above-the-limb events. The proton track geometry is characterized by its viewing angle (i.e., the off-nadir angle at the position of the detector) as described in Sec.~\ref{sec:atmosphere}. The vertical dotted lines in both panels correspond the viewing angle of the Earth's limb at the respective altitude.}
    \label{fig:nphotons_altitude_comparison}
\end{figure}

\begin{figure}[htb]
    \begin{minipage}{\figcolwidth}
    \centering
    \includegraphics[width=\linewidth]{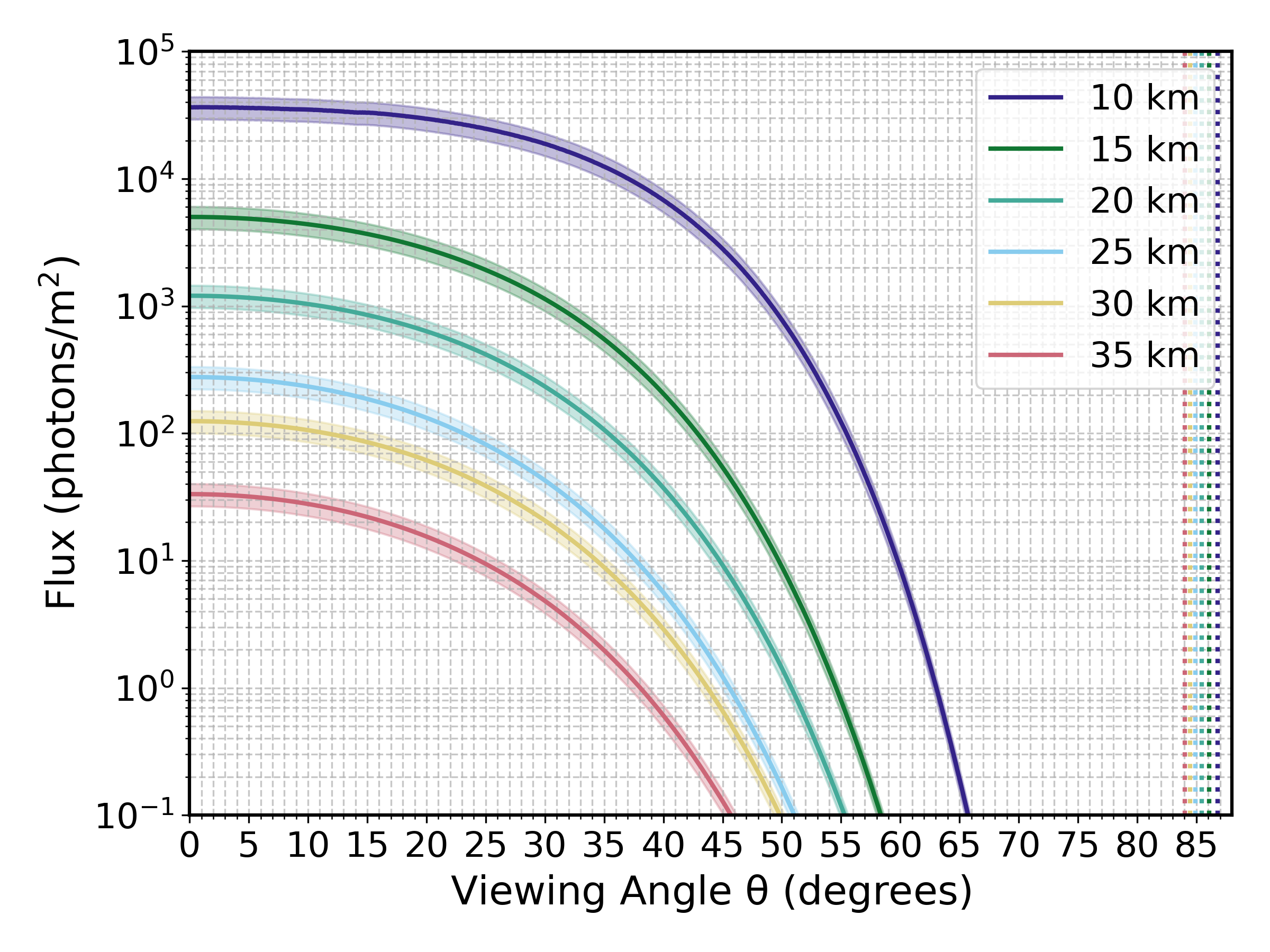}\\
    (a)
    \end{minipage}\hfill
    \begin{minipage}{\figcolwidth}
    \centering
    \includegraphics[width=\linewidth]{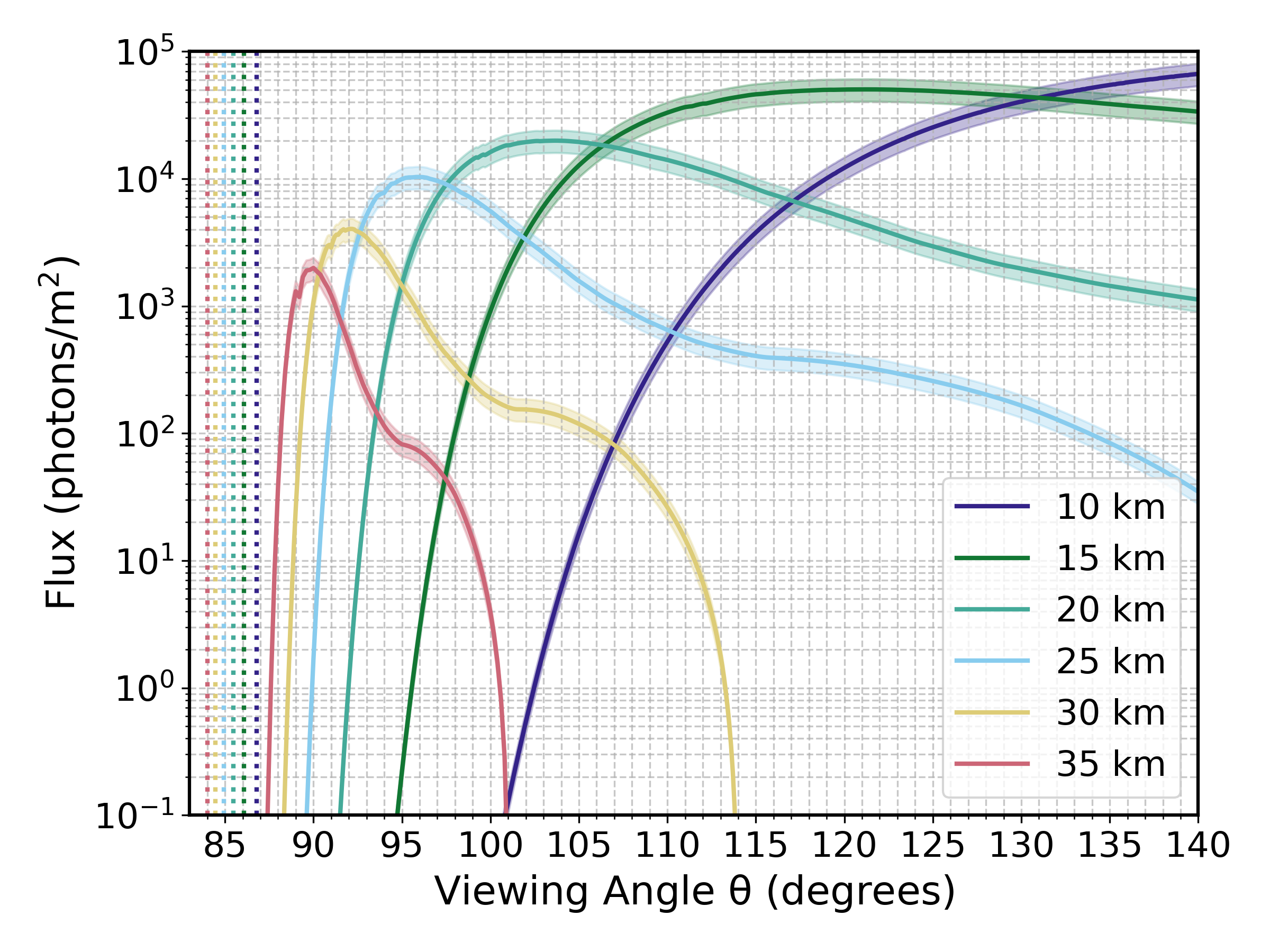}\\
    (b)
    \end{minipage}
    \caption{(a) Effective X-ray flux (from \qtyrange{10}{E4}{keV}) at the position of a detector at several altitudes for below-the-limb events from \qty{100}{PeV} proton-induced EAS. (b) The same, but for above-the-limb events. The proton track geometry is characterized by its viewing angle (i.e., the off-nadir angle at the position of the detector) as described in Sec.~\ref{sec:atmosphere}. The vertical dotted lines in both panels correspond the viewing angle of the Earth's limb at the respective altitude.}
    \label{fig:fluxes_altitude_comparison}
\end{figure}

The X-ray photon fluxes at the position of the detector were computed by solving Eq.~\ref{eq:master_formula} numerically and estimating the footprint area following the procedure in Sec.~\ref{sec:areas}. Figure~\ref{fig:nphotons_altitude_comparison} shows the absolute number of photons arriving at the detector position for below- and above-the-limb events for a set of detector altitudes between \qtyrange{10}{35}{km} from a \qty{100}{PeV} primary assuming a first interaction depth of \qty{30}{g cm^{-2}} for the former, and a depth equivalent to the expected decay length of a $\tau$ lepton for the latter (fixed at $\sim \qty{5}{km}$ for a \qty{100}{PeV} $\tau$). Figure~\ref{fig:fluxes_altitude_comparison} shows the photon fluxes for the same conditions. The error bands shown were obtained by considering a conservative \qty{20}{\percent} uncertainty on the number of electrons in the shower obtained from the Nerling energy distribution.

As can be seen from Fig.~\ref{fig:nphotons_altitude_comparison}, for the same observation altitude, the number of photons arriving to the detector is about an order of magnitude smaller for below-the-limb events. The reason for this is two-fold. First, below-the-limb events come from showers initiated by leptons emerging from the Earth and propagating upwards. With respect to horizontal showers, up-going showers develop in denser layers of the atmosphere, and this is particularly true for the earlier stages of development when X-ray synchrotron emission is most relevant. As a result, the shower develops over shorter lengths, and consequently, over shorter timescales. Since the power emitted in synchrotron radiation by a single particle is constant in time, shorter emission timescales lead directly to fewer emitted photons. Second, since the shower develops in a denser atmosphere, the effect of atmospheric absorption is more important.

From Fig.~\ref{fig:nphotons_altitude_comparison}, it can also be seen that the number of photons on the detection plane decreases as the altitude decreases. For below-the-limb events, this is due to the fact that as the altitude decreases, the detector gets closer to the (fixed) point of the $\tau$ lepton decay: the shower thus develops less before arriving to the the detector and the photon emission is smaller. On the other hand, for above-the-limb events, as the altitude decreases, near-horizontal showers develop in a layer of the atmosphere that is increasingly less rarefied. Consequently, photon emission is smaller due to both a shorter timescale of shower development and larger atmospheric photon absorption. Further, as the altitude decreases, the viewing angle at which photon emission becomes appreciable both increases and gets farther away from the Earth's limb viewing angle due to the fact that the grammage of trajectories close to the limb increases, and photon emission is entirely absorbed before reaching the detector.

Lastly, by comparing Fig.~\ref{fig:nphotons_altitude_comparison} and Fig.~\ref{fig:fluxes_altitude_comparison} it becomes clear that as the altitude decreases, so does the area of the photon footprint at the position of the detector. This is primarily due to the fact that the shower increasingly develops in a denser atmosphere, as discussed before. Since the lateral scale of the electrons in the shower depends on the local atmospheric density (through its Molière radius, or through a scaled version of it as discussed in Sec.~\ref{sec:areas}), as the density decreases so does the lateral scale. Consequently, the shower electrons are distributed over smaller areas and the resulting photon footprint areas are also smaller. These footprint areas can be as large as $\mathcal{O}(\qty{100}{m^2})$ at \qty{35}{km}, and reach sub-\unit{m^2} level at \qty{10}{km}. The latter scenario presents a serious challenge for detection since it largely constraints the phase space of primary candidates that can produce a photon footprint that intersects a given detection area (as will be discussed in detail in the following section).

\section{Detector Acceptance}\label{sec:acceptance}

As is commonly known, the event rate $R$ observed by a detector is given by the convolution of its acceptance (or geometric factor, depending on convention) with the incoming particle flux~\cite{sullivan1971geometric}. For an isotropic flux, such as that of CRs, the rate is expressed as,
\begin{equation}\label{eq:rate_definition}
    R = \int_{E_\text{min}}^\infty d E \, G(E) \, \phi (E)
\end{equation}
where ${E_\text{min}}$ is the primary threshold energy below which the detector efficiency is zero, $G(E)$ is the acceptance, and $\phi(E)$ is the isotropic CR flux. Eq.~\ref{eq:rate_definition} thus defines the acceptance as the proportionality factor between the incident flux and the observed counting rate. Mathematically, this factor can be written as an integral of the detector efficiency over its entire collecting area and over all angles, as follows,
\begin{equation}\label{eq:geometric_factor_analytic}
	G(E)= \int_S d \sigma \int_\Omega d\omega  \; (\hat{\sigma} \cdot \hat{r}) (\omega) \, \epsilon (E, \vec{\sigma}, \omega)
\end{equation}
where $d \sigma$ is an infinitesimal area element of the detector collecting area with normal $\hat{\sigma}$, $d \omega$ an element of solid angle, $\hat{r}$ the unit vector pointing in the direction of $\omega$, and $\epsilon(E, \vec{\sigma}, \omega)$ the detector efficiency for a given energy, surface element, and incoming direction.

For simple detector geometries, Eq.~\ref{eq:geometric_factor_analytic} may be solved analytically, or numerically. However, the problem becomes intractable rather quickly for more complex geometries. A standard approach in these cases is to compute the acceptance using the Monte Carlo (MC) technique. This method defines a candidate generation surface (that either contains or envelops the detector area), and samples candidate events with uniformly distributed starting positions and isotropically distributed directions~\cite{crannell_MC, sullivan1971geometric}. Using this procedure, the acceptance is obtained as:
\begin{equation}\label{eq:geometric_factor_mc}
	G (E)= \frac{N_\text{det}}{N_\text{gen}} \times (A_\text{gen} \, \Omega_\text{gen})
\end{equation}
where $N_\text{gen}$ is the number of simulated primary candidates (with energy $E$), $N_\text{det}$ the number that satisfy the detection criteria, and $A_\text{gen}$ and $\Omega_\text{gen}$ the total generation area and solid angle spanned by that area in the detector's coordinate frame, respectively.

In the following, we outline the procedure adopted to estimate the acceptance and event rate for a simplified detector configuration at several altitudes, using a modified version of the MC technique, for computational efficiency. In particular, when the generation area is subdivided into equal-area cells and a sufficiently large $N_\text{gen}$ is used, each cell will contain the same number of generated candidates due to the uniform sampling. Introducing $\xi_\text{i}(E)$ as the proportionality factor between the number of generated candidates with energy $E$ in the i-th cell, $N_\text{gen,i} \,(E)$, and the number that are detected, $N_\text{gen,i} \,(E)$, Eq.~\ref{eq:geometric_factor_mc} can be rewritten as:
\begin{equation}\label{eq:geometric_factor_mc_there}
\begin{split}
	G(E) &= (A \, \Omega)_\text{gen} \times \sum_i \frac{N_\text{det,i}}{N_\text{gen}} = (A \, \Omega)_\text{gen} \times \sum_i \frac{N_\text{det,i} \, \xi_i(E)}{N_\text{gen}} \\
         &=  (A \, \Omega)_\text{gen} \times \frac{1}{N_\text{cells}} \sum_i \; \xi_i(E)
\end{split}
\end{equation}
where the last equality follows from the uniform sampling and equal-area subdivision. In Sec.~\ref{sec:bootstrap} we describe the MC procedure used to compute the $\xi_i(E)$ factors, and in Sec.~\ref{sec:results} we present our estimates of the acceptance and event rate for a circular detector with  radius of \qty{1}{m} and a \qty{70}{\degree} half-aperture field-of-view (FoV), oriented towards the Earth's limb at altitudes between \qtyrange{20}{30}{km}.

\subsection{MC Procedure}\label{sec:bootstrap}

To compute the acceptance according to Eq.~\ref{eq:geometric_factor_mc_there}, we first define a candidate generation surface corresponding to a spherical band at a radius equal to the Earth's radius, $R_\text{E}$, plus a height of \qty{112.79}{km}, which we take as the effective upper boundary of the atmosphere (compatible with the US standard atmosphere model~\cite{atmosphere1976us}). The azimuth and zenith bounds of this band are established using the detector's orientation, its field-of-view, and the range of angles over which at least one photon arrives at the position of the detector at the highest simulated primary energies (cf. Figs.~\ref{fig:nphotons_altitude_comparison} and~\ref{fig:fluxes_altitude_comparison}). This band is then divided into $N_\text{cells}$ equal-area cells. The sum of all the cell areas is the total generation area, $A_\text{gen}$, whereas the solid angle, $\Omega_\text{gen}$, is computed as the solid angle spanned by this band from the detector position. The $\xi_\text{i}$ factors in Eq.~\ref{eq:geometric_factor_mc_there} (hereafter, the survival probabilities) are then computed through a bootstrap-based procedure described in the following.

\begin{figure}[htb]
    \begin{minipage}{\figcolwidth}
    \centering
    \includegraphics[width=\linewidth]{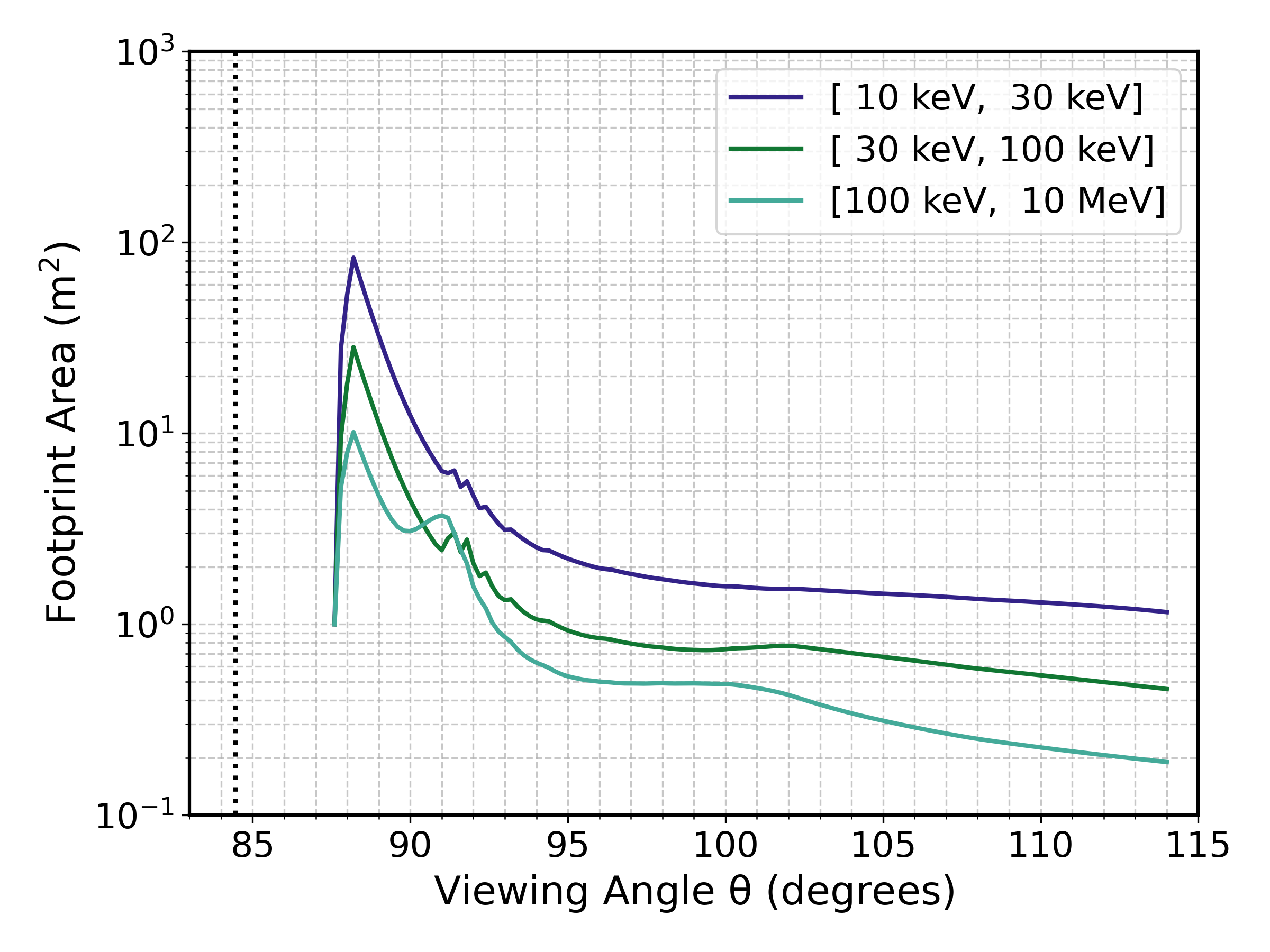}\\
    (a)
    \end{minipage}\hfill
    \begin{minipage}{\figcolwidth}
    \centering
    \includegraphics[width=\linewidth]{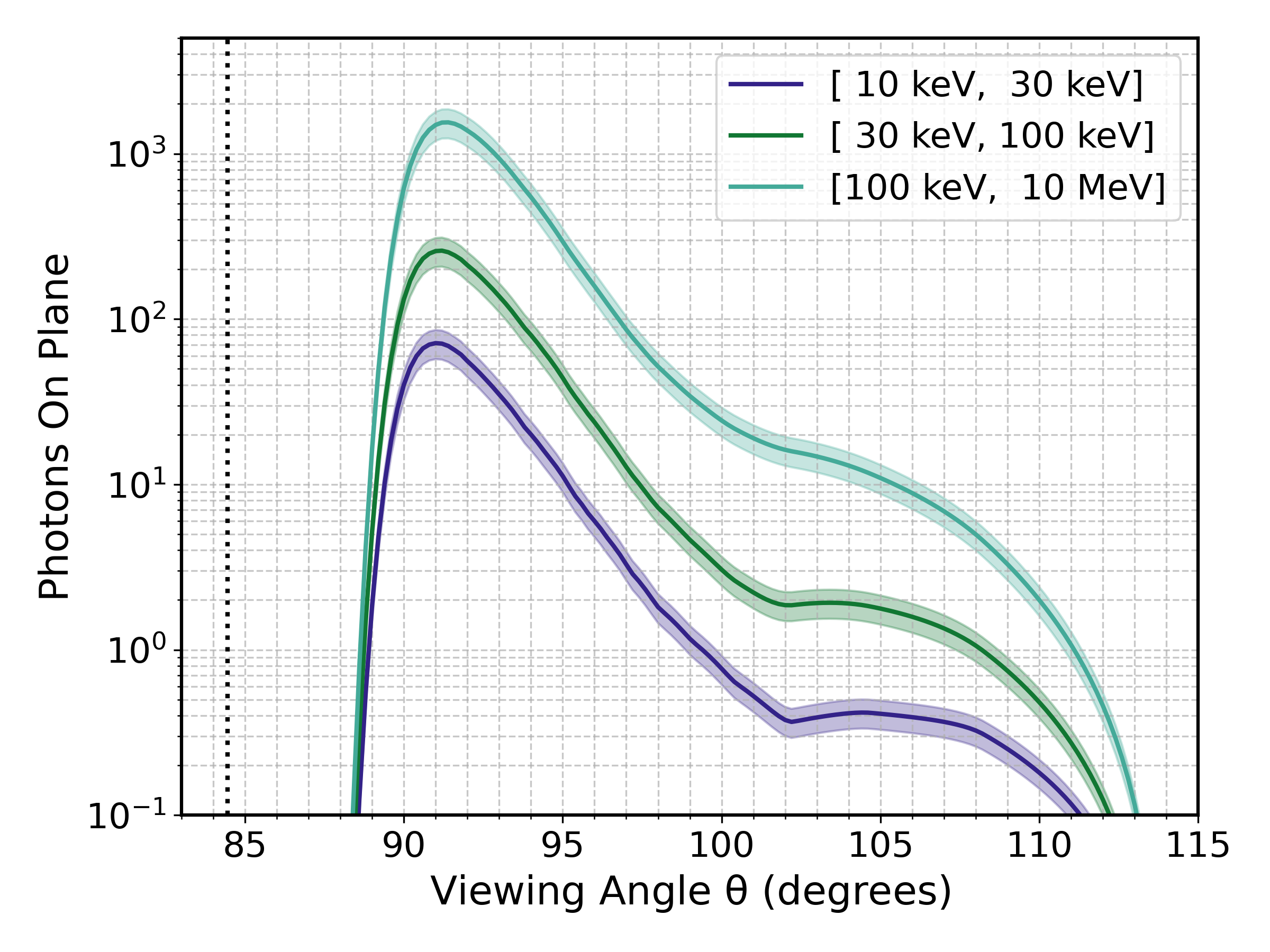}\\
    (b)
    \end{minipage}
    \caption{(a) Example of the illuminated area on the shower plane at the position of the detector differentiated in terms of the photon energy band for a detector at \qty{30}{km}. The peak structure of the area curves between \qty{90}{\degree} and \qty{95}{\degree} are an effect from the geomagnetic deviation of electrons in the shower. (b) Example of the number of photons arriving to the detector differentiated in terms of the photon energy band.}
    \label{fig:area_and_nphotons}
\end{figure}

For a given detector altitude, we first perform a scan over CR energies and viewing angles to precompute two key quantities: the number of photons reaching the detection plane, and the illuminated footprint area of these photons on that plane. Further, these quantities were computed for three different photon energy bands: a \enquote{soft} band (\qtyrange{10}{30}{keV}), a \enquote{hard} band (\qtyrange{30}{100}{keV}), and a \enquote{gamma} band (\qtyrange{100}{E4}{keV}). All of these quantities are obtained as a function of the primary energy and viewing angle using the procedure described in Sec.~\ref{sec:computational_approach} and Sec.~\ref{sec:fluxes}. Figure~\ref{fig:area_and_nphotons} provides an example of the precomputed footprint areas and number of transmitted photons for a \qty{10}{PeV} proton primary and a detector at \qty{30}{km} altitude. These quantities are computed for a single direction, corresponding to a primary traveling along the line connecting its point of origin on the sky to the detector center at fixed altitude.

\begin{figure}[htb]
    \centering
    \includegraphics[width=0.49\linewidth]{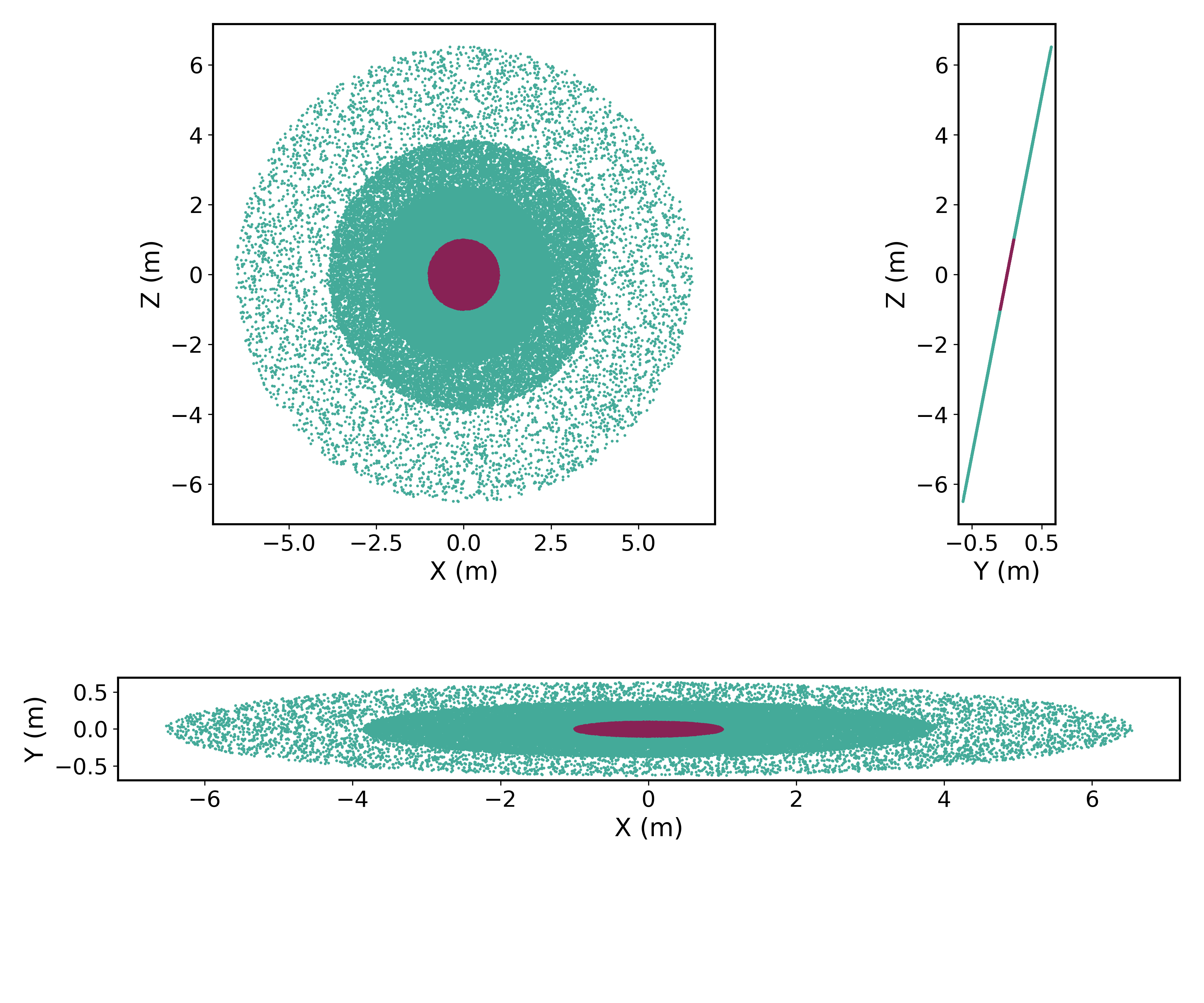}
    \includegraphics[width=0.49\linewidth]{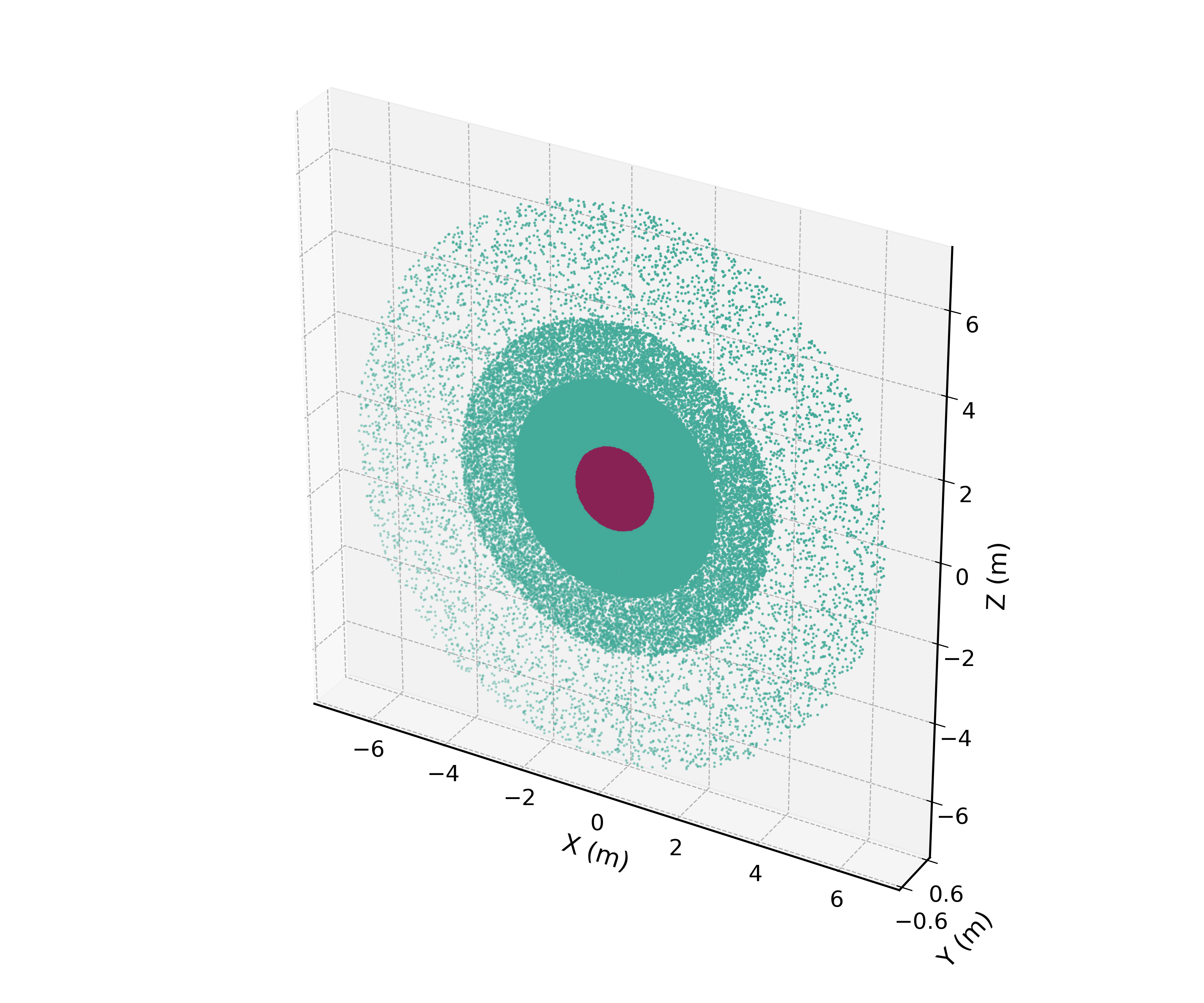}
    \caption{Example of a simulated photon footprint on the detection plane for \num{E5} photons. The detector geometry is a disk with \qty{1}{m} radius as described in the text. Magenta points fall within the detector geometry, while teal points lie outside of it. The hit proportion is obtained as the ratio between these counts, estimated through bootstrapping (see text).}
    \label{fig:projected_footprint}
\end{figure}

For every cell center on the generation surface, we draw a line connecting that point to the detector center. Around this reference direction, we define a grid of possible candidate directions represented by small angular deviations ($\Delta \theta'$, $\Delta \phi'$). For each direction in this grid (at a fixed point in the sky) we generate a set of photon positions distributed on a \enquote{shower plane} (i.e., a plane perpendicular to the reference direction vector) and project these points onto the detector plane. The generated photon position are distributed on three concentric circles, one for each photon energy band, and the ratio of generated positions on every circle is determined from the ratio of the number of photons arriving at the position of the detector between any two photon energy bands (see Fig.~\ref{fig:projected_footprint}). A photon hit proportion is then computed as the ratio between the number of points falling within the detector geometry (the magenta points in Fig.~\ref{fig:projected_footprint}) and the total number of generated points (the teal points in Fig.~\ref{fig:projected_footprint}). To improve the statistical robustness of this hit proportion estimate, we apply a bootstrap procedure, namely: the projected positions are resampled with replacement, the hit-to-miss ratio is recomputed for each sample, and the resulting distribution of the hit proportion estimator is used to derive its median and a \qty{95}{\percent} confidence interval. 

\begin{figure}[htb]
    \begin{minipage}{\figcolwidth}
    \centering
    \includegraphics[width=\linewidth]{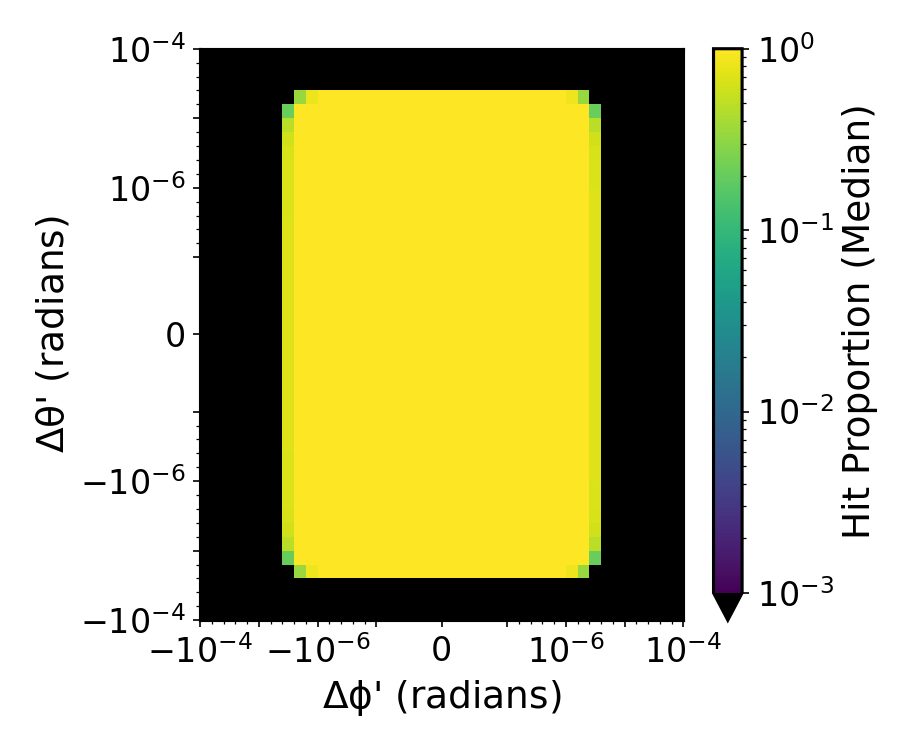}\\
    (a)
    \end{minipage}\hfill
    \begin{minipage}{\figcolwidth}
    \centering
    \includegraphics[width=\linewidth]{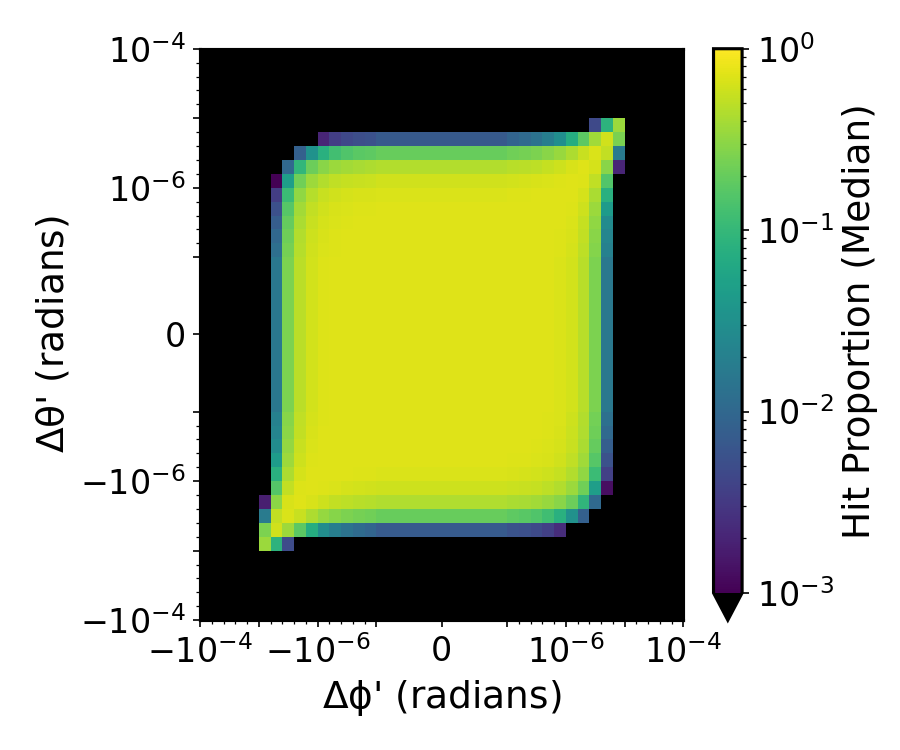}\\
    (b)
    \end{minipage}
    \caption{(a) Map of the photon hit proportions for a detector at \qty{20}{km}, as defined in the text, for primaries originating at a sky position corresponding to a zenith angle of \qty{24.53}{\degree}, and an azimuth angle \qty{90}{\degree} as measured from the position of the detector. (b) The same, but for primaries originating from a sky position corresponding to a zenith angle of \qty{24.53}{\degree}, and an azimuth angle of \qty{20}{\degree}. The detector geometry use to compute these maps is that shown in Fig.~\ref{fig:projected_footprint}. In both panels, the elongation of the area along $\Delta \theta'$ on the grid where the hit proportion is larger than zero is due to the effect of projecting the circular photon footprint onto the detection plane at a slant zenith angle, where the area is transformed into an ellipse. On panel (b), the footprint arrives deviated both in zenith and azimuth relative to the detection plane normal, leading to the observed distortions.}
    \label{fig:hit_probabilities}
\end{figure}

This bootstrap procedure is repeated for all directions in the ($\Delta \theta'$, $\Delta \phi'$) grid and for every cell on the generation surface. After this stage, each cell midpoint on the sky is associated with a map of hit proportion estimators, $\hat{p}_\text{m,n}$ with $m$ and $n$ the indices of the 2D direction grid, corresponding to all isotropic directions from that point. Figure~\ref{fig:hit_probabilities} shows two examples of the resulting hit proportion maps for a detector at \qty{20}{km} altitude and for two sky positions with identical zenith angles but different azimuths. To obtain  single survival probability $\xi_i$ for every cell on the generation surface, two factors must be considered. The first is the probability that a candidate travels in a given direction of the grid, which is given by the ratio between the solid angle element of the direction grid cell, $\omega_\text{m,n}'$, and the full solid angle $4 \pi$. The second is the probability that the detector triggers, given a total number of photons on the plane, $N_\mathrm{\gamma}$, and the hit proportion, $\hat{p}_\text{m,n}$, for a threshold of at least $N_\text{thr}$ photons within the detector geometry. This probability is given by the complement of the cumulative distribution function (CDF) of the binomial distribution:
\begin{equation}\label{eq:binomial}
    P(N_\text{hit} \ge N_\text{thr})_\text{m, n} = 1 - F(N_\text{thr}; N_\mathrm{\gamma}, \hat{p}_\text{m,n})
\end{equation}
where $F(N_\text{thr}; N_\mathrm{\gamma}, p_\text{m,n})$ is the binomial CDF.

Given that the number of photons arriving at the detector position, $N_\mathrm{\gamma}$, depends on both geometry and primary energy, the survival probabilities (and therefore the acceptance) are computed as functions of energy. Combining all terms described above, the survival probability for  fixed position in the sky is given by:
\begin{equation}
    \xi_\text{i} (E) = \sum_{m,n}  \text{P}(N_\text{hit} \ge N_\text{thr}; N_\mathrm{\gamma,i} (E), \hat{p}_{m,n}) \times \left( \frac{\omega_{m,n}'}{4 \pi} \right)
\end{equation}
where $m, n$ are the indices of the cell on the direction grid for a fixed point on the sky, and $i$ is the index identifying the equal-area cells of the generation surface. From the expression above, the detector acceptance and event rate follow from Equations~\ref{eq:geometric_factor_mc_there} and~\ref{eq:rate_definition}, respectively.

\subsection{Results}\label{sec:results}

\begin{figure}[htb]
    \begin{minipage}{\figcolwidth}
    \centering
    \includegraphics[width=\linewidth]{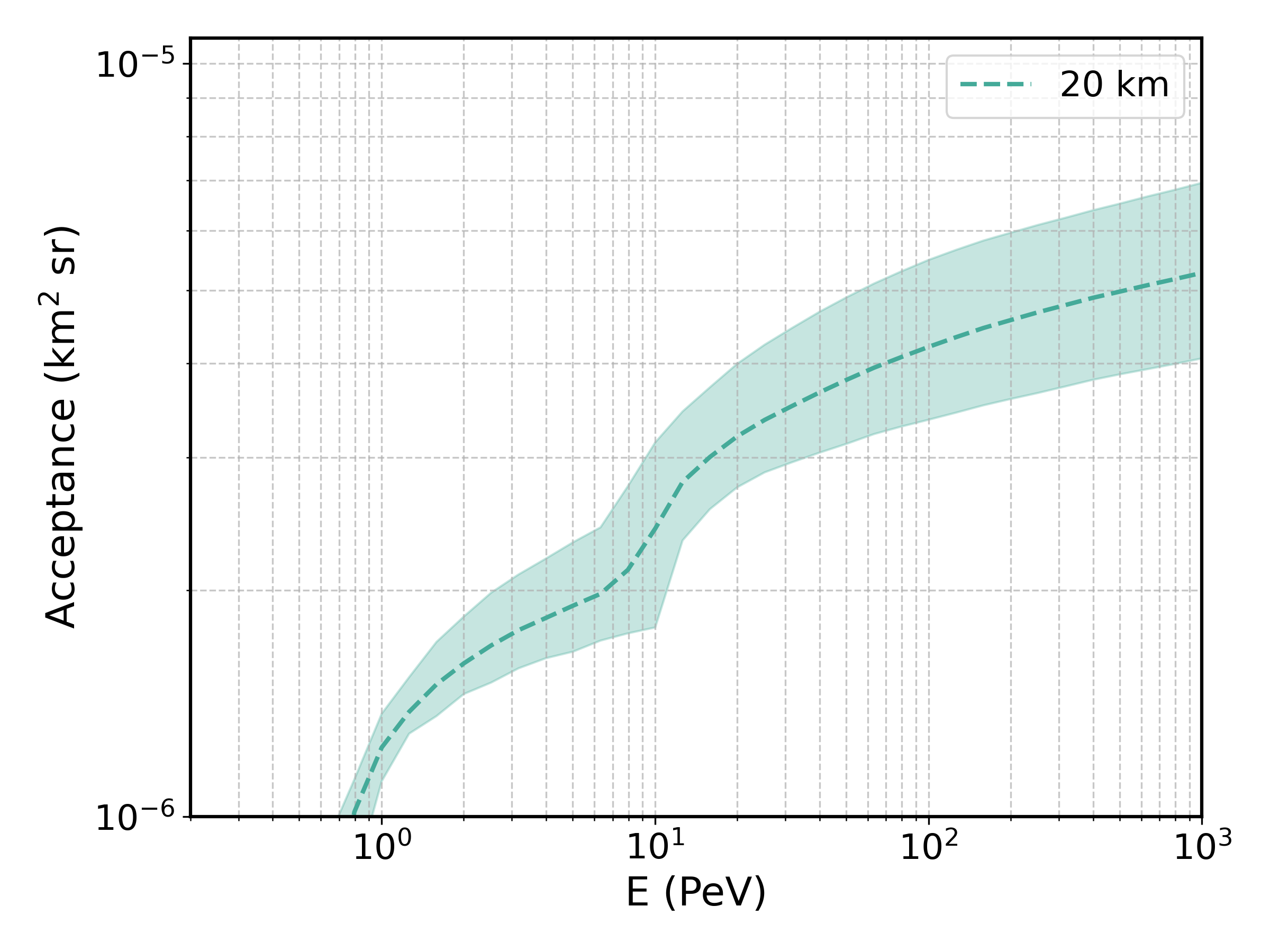}\\
    (a)
    \end{minipage}\hfill
    \begin{minipage}{\figcolwidth}
    \centering
    \includegraphics[width=\linewidth]{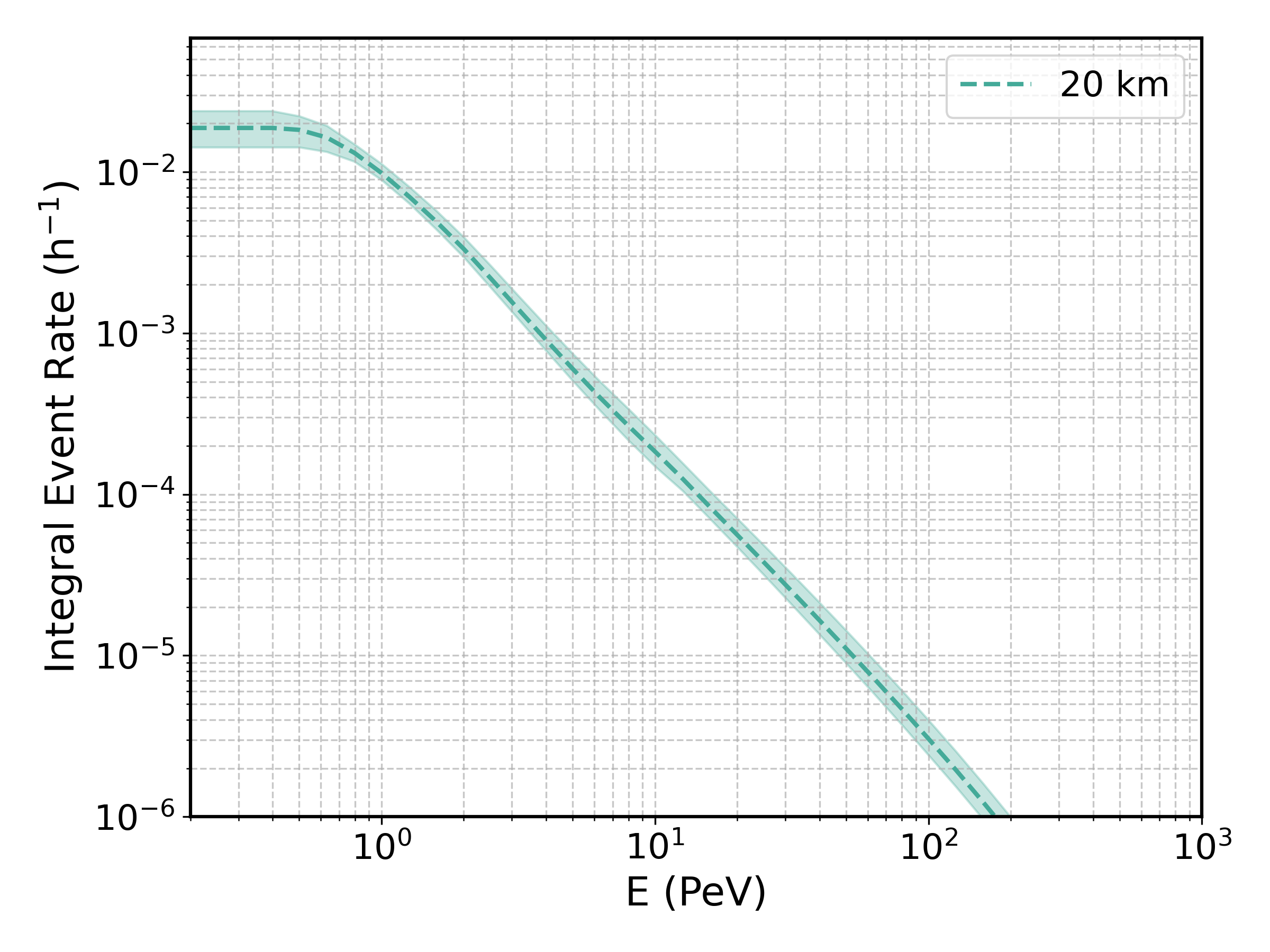}\\
    (b)
    \end{minipage}
    \caption{(a) Detector acceptance, and (b) integral event rate for a detector at \qty{20}{km} with a geometry as described in the text. The dotted line shows the estimated mean value and the error band was obtained by bracketing the uncertainties from the shower distributions (see text for details).}
    \label{fig:results_20km}
\end{figure}

\begin{figure}[htb]
    \begin{minipage}{\figcolwidth}
    \centering
    \includegraphics[width=\linewidth]{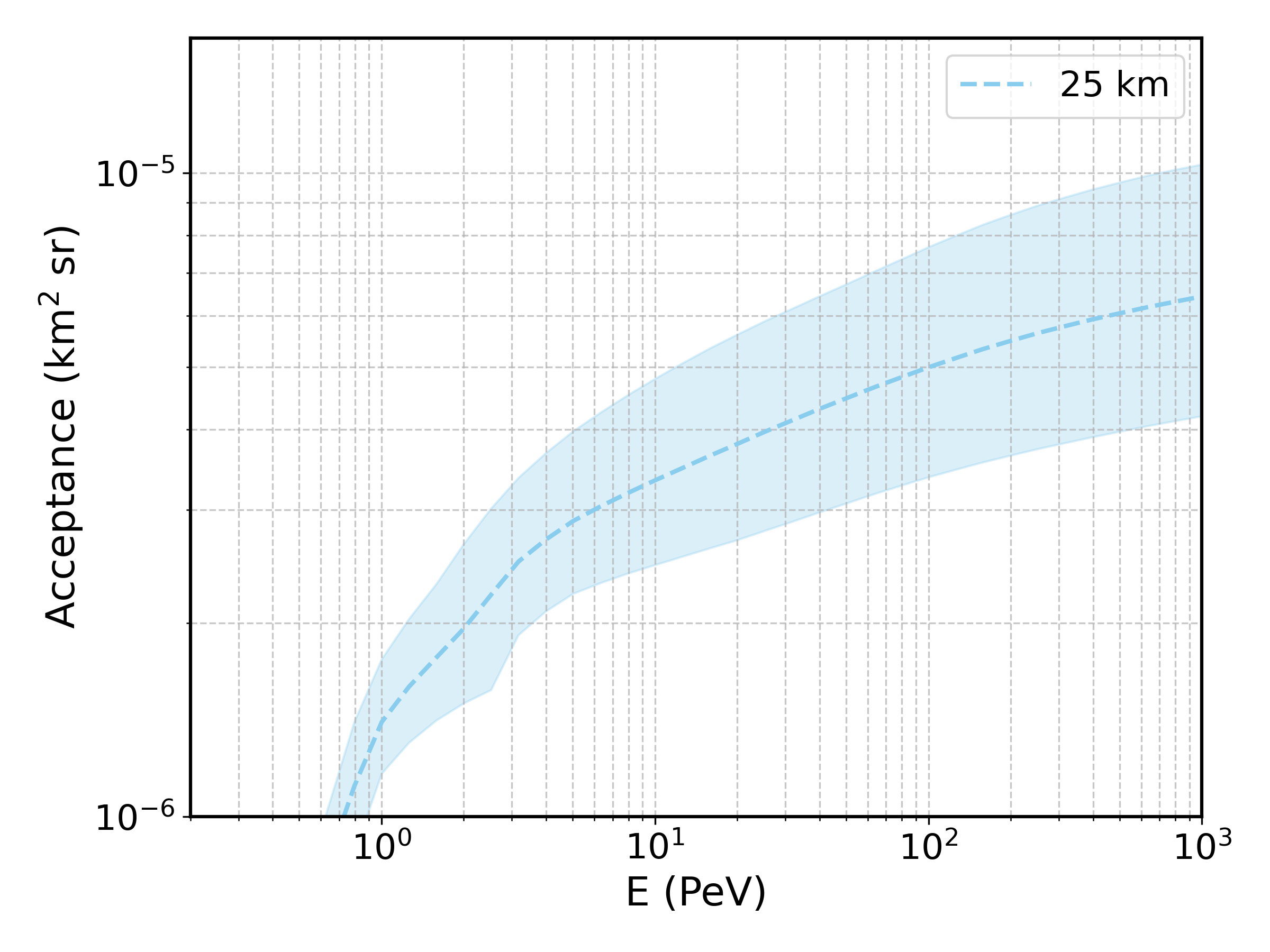}\\
    (a)
    \end{minipage}\hfill
    \begin{minipage}{\figcolwidth}
    \centering
    \includegraphics[width=\linewidth]{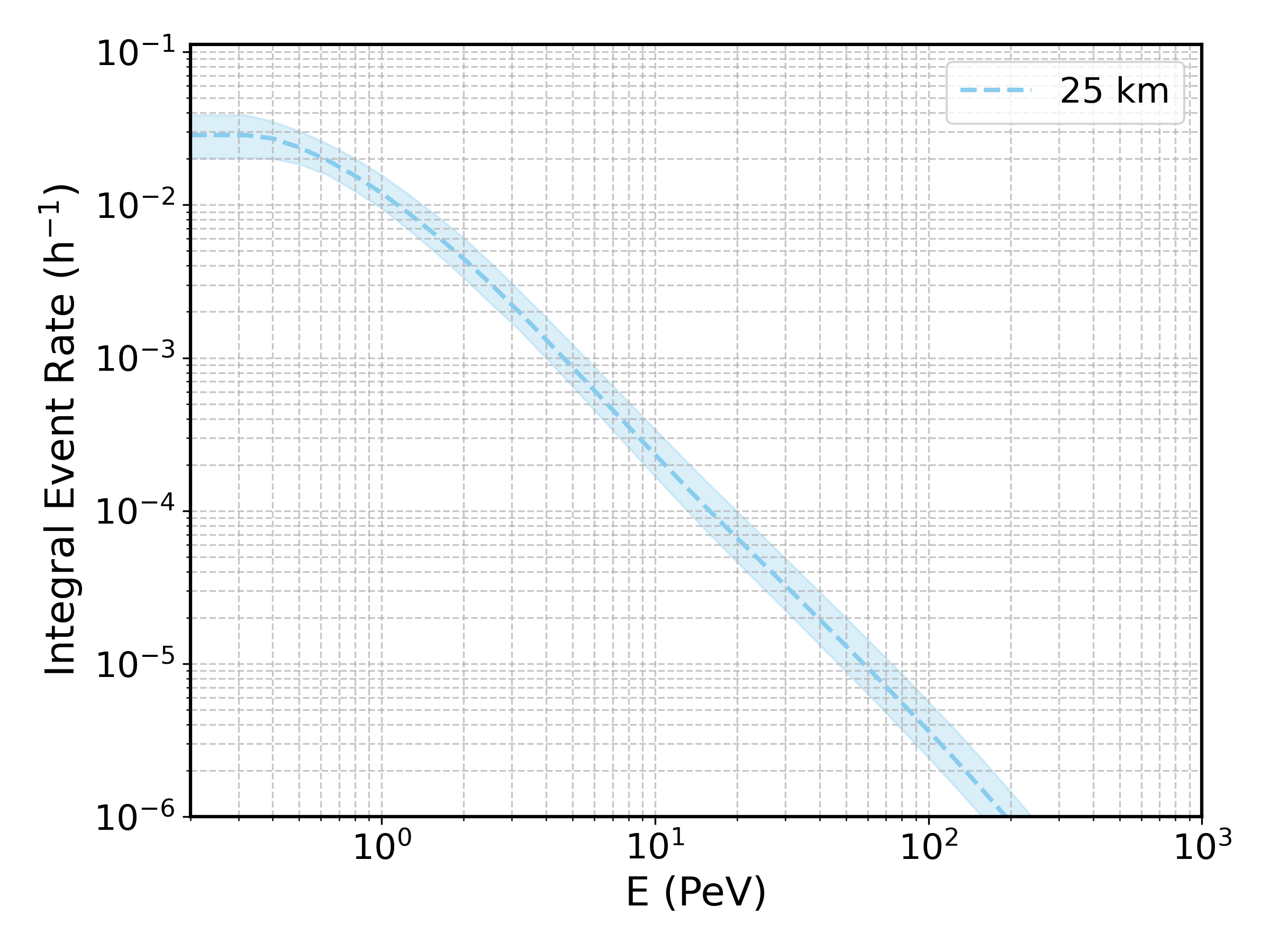}\\
    (b)
    \end{minipage}
    \caption{(a) Detector acceptance, and (b) integral event rate for a detector at \qty{25}{km} with a geometry as described in the text. The dotted line shows the estimated mean value and the error band was obtained by bracketing the uncertainties from the shower distributions (see text for details).}
    \label{fig:results_25km}
\end{figure}

\begin{figure}[htb]
    \begin{minipage}{\figcolwidth}
    \centering
    \includegraphics[width=\linewidth]{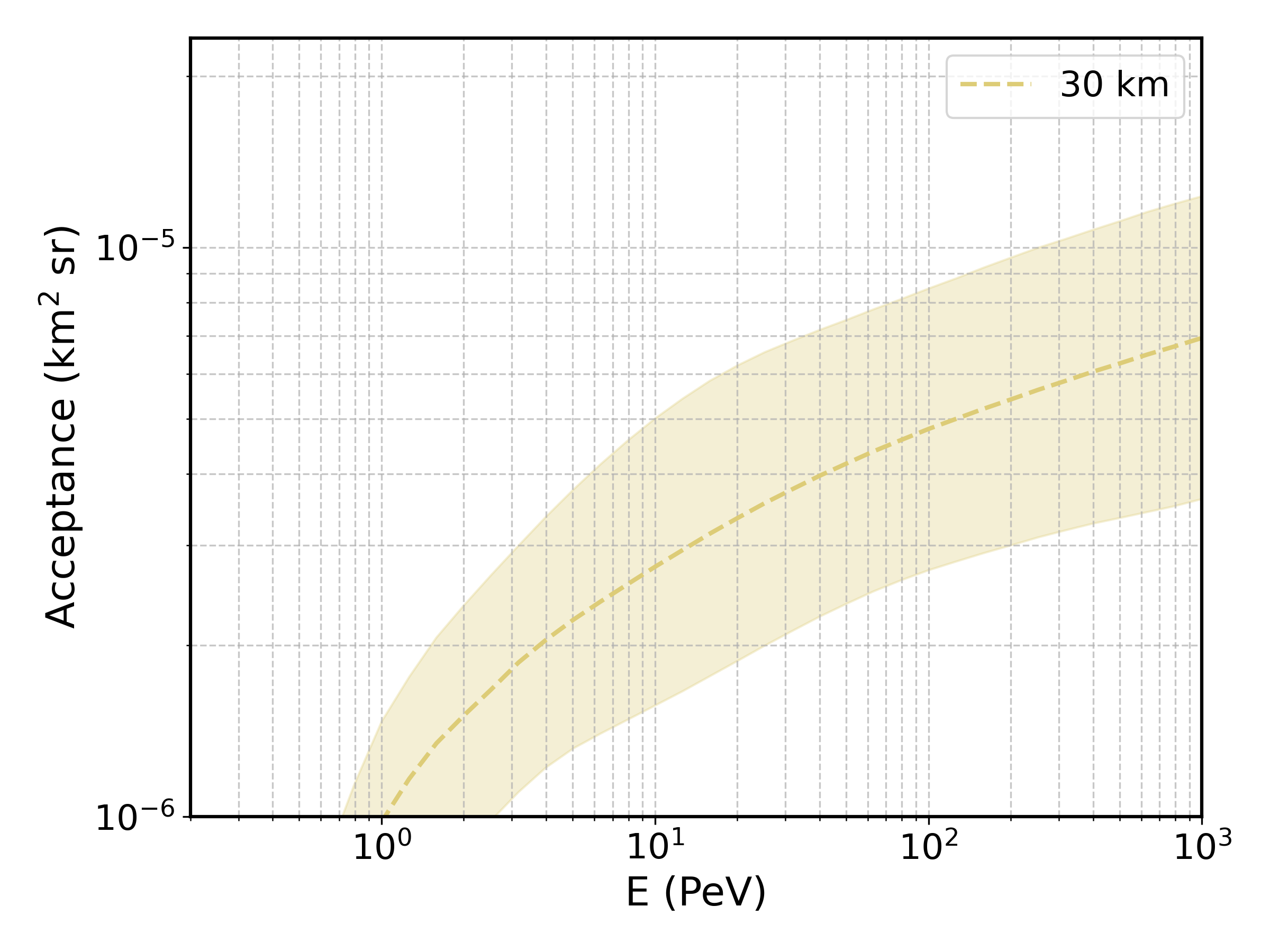}\\
    (a)
    \end{minipage}\hfill
    \begin{minipage}{\figcolwidth}
    \centering
    \includegraphics[width=\linewidth]{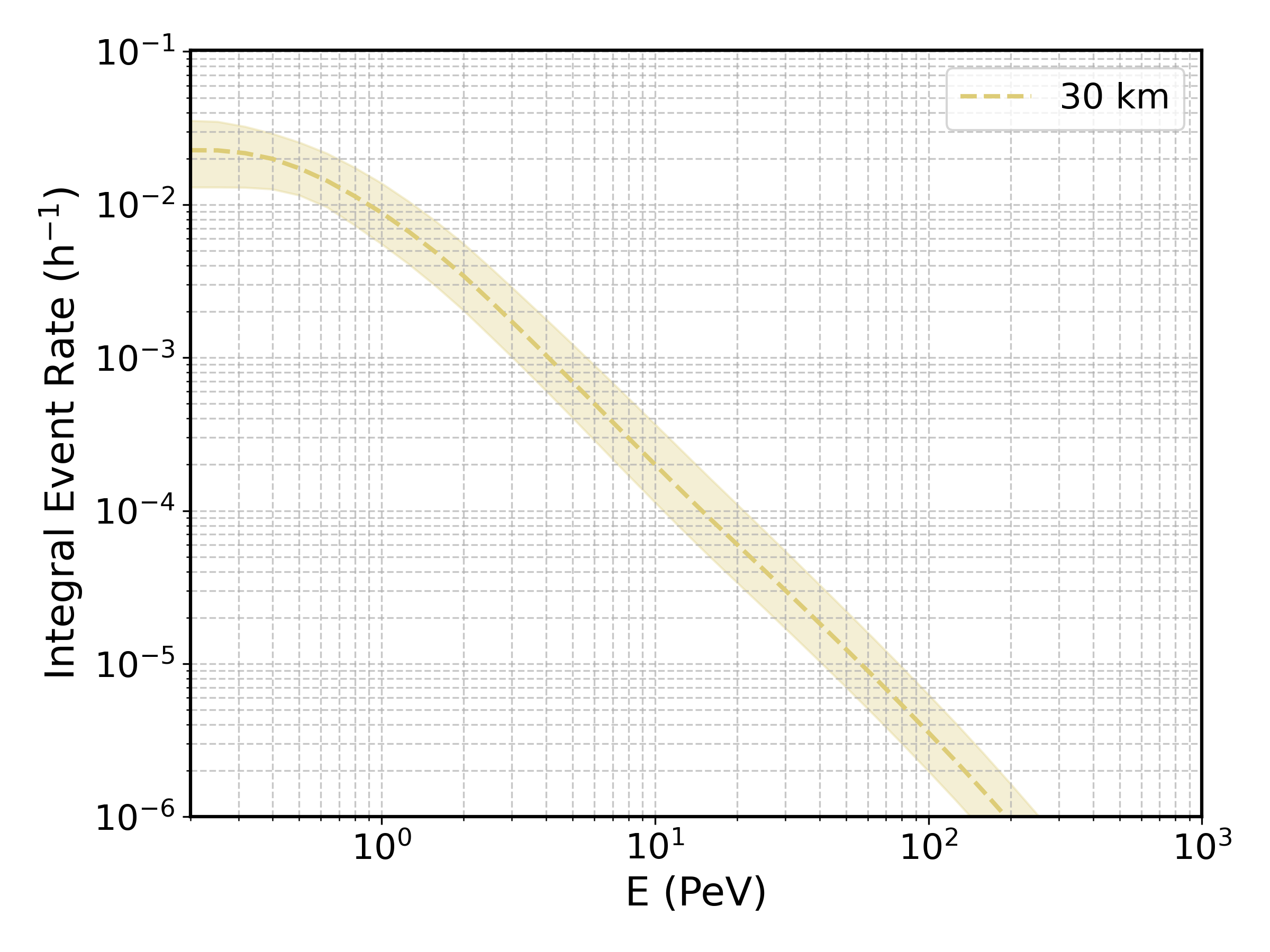}\\
    (b)
    \end{minipage}
    \caption{(a) Detector acceptance, and (b) integral event rate for a detector at \qty{30}{km} with a geometry as described in the text. The dotted line shows the estimated mean value and the error band was obtained by bracketing the uncertainties from the shower distributions (see text for details).}
    \label{fig:results_30km}
\end{figure}

The acceptance and event rate of a circular detector with  radius of \qty{1}{m} and a \qty{70}{\degree} half-aperture FoV, oriented towards the Earth's limb was computed using the bootstrap MC procedure described in Sec.~\ref{sec:bootstrap} for three flight altitudes: \qty{20}{km}, \qty{25}{km}, and \qty{30}{km}. These three altitudes were chosen since a preliminary altitude scan showed this altitude range to be ideal for the observation EAS X-ray synchrotron signal. Figures~\ref{fig:results_20km}, \ref{fig:results_25km} and \ref{fig:results_30km} show the results obtained for \qty{20}{km}, \qty{25}{km}, and \qty{30}{km} respectively. The error band was obtained by considering a conservative \qty{20}{\percent} systematic uncertainty on the number of photons arriving to the plane (coming from the Nerling electron energy distribution) and by bracketing the variation on the computed value for the acceptance from considering different scales of the shower lateral distribution (cf. Sec.~\ref{sec:electron_ldf}).

The results show that for the altitude range under consideration, the detector acceptance is on the order of $\sim \qty{E-6}{km^2 sr}$ (\qty{1}{m^2 sr}). This value is consistent with what would be expected from a first order estimate of the acceptance of a two element telescope given the scale of the detector area, $\mathcal{O}(\qty{1}{m^2})$, the area of the aperture (which corresponds to the area of the generation spherical band), $\mathcal{O}(\qty{E6}{km^2})$, and the typical distances between the points on the generation surface and the detector position, $\mathcal{O}(\qty{E3}{km})$. This is due to the fact that the area of the photon footprint is typically on the order of \qtyrange{1}{10}{m^2}, which is comparable to the area of the detector. Given this consideration, the condition for detecting an EAS through its X-ray synchrotron emission translates roughly to requiring the shower-axis intersect the area of the detector.

Given the integral rates shown in Fig.~\ref{fig:results_20km}, Fig.~\ref{fig:results_25km}, and Fig.~\ref{fig:results_30km} we conclude that a balloon mission carrying a detector payload with a geometry similar to that used for this study could hope to observe $\mathcal{O}(10)$ events during a month's flight with a full duty cycle. Specifically, at \qty{20}{km} the best mean estimate for the number of events is \num{12.2} with a \qty{95}{\percent} confidence interval (CI) of [\num{7.97}, \num{17.0}]. At \qty{30}{km}, the best estimate is \num{14.6} ([\num{8.35}, \num{22.6}] \qty{95}{\percent} CI). Finally, at \qty{25}{km}, the best estimate is \num{18.5} ([\num{13.0}, \num{24.9}] \qty{95}{\percent} CI).

\section{Conclusions}\label{sec:conclusions}

In this work, we have developed a novel simulation framework to estimate X-ray geo-synchrotron emissions from EAS. Following the computational philosophy of~\cite{cummings2021modeling1, cummings2021modeling2}, we applied the theory of incoherent synchrotron emission to the case of EAS to derive a differential equation for the number of photons arriving at the plane of a sub-orbital detector (cf. Sec.~\ref{sec:computational_approach}), in particular Eq.~\ref{eq:master_formula}. We solved this equation numerically using results from \texttt{CORSIKA} simulations and parameterizations of the shower-electron distributions. These parameterizations were also used to estimate the footprint (i.e., the photon spatial distribution) in the detection plane, enabling the calculation of expected fluxes as functions of altitude and viewing angle (cf. Sec.~\ref{sec:fluxes}).

We have found that the number of detected photons increases with altitude as a result of both the longer shower development timescales and the reduced atmospheric attenuation in the rarefied upper atmosphere (cf. Sec.~\ref{sec:fluxes_results}). Conversely, the range of observing angles that provide sufficient grammage for shower development decreases with altitude, as one approaches the upper boundary of the atmosphere. These trends yield an optimal observation altitude range of approximately \qtyrange{20}{30}{km}. Finally, we have developed a bootstrap Monte Carlo procedure that uses the results of the numerical simulations to estimate the acceptance and event rate of simplified detector geometries (cf. Sec.~\ref{sec:acceptance}). For a \qty{1}{m} radius, \qty{70}{\degree} half-aperture circular detector observing the Earth's limb from altitudes in the \qtyrange{20}{30}{km} range, we have found detector acceptances at the $\mathcal{O}(\qty{1}{m^2 sr})$ level and integral event rates at the $\mathcal{O}(10)$ per month level. These results suggest that the X-ray geo-synchrotron channel can constitute a viable and complementary method for PeV-scale indirect CR observation from sub-orbital platforms, particularly employing hybrid detection techniques.

We emphasize that the results presented here are a first step toward characterizing X-ray geo-synchrotron emission. As discussed throughout this work, the emission is dominated by very young showers ($s \lesssim 0.5$), where the principle of shower universality becomes questionable and where little work has been done to characterize the properties of high-energy electrons. Planned follow-on studies will include refinements of the methodology used to compute the spatial distribution of electrons in the detection plane, improvements in the Monte Carlo acceptance procedure, and high-statistics, detailed simulations of electron distributions in young showers.

Beyond EAS, we note that X-ray emission may also arise from other physical processes. In particular, \cite{ling_background} describes an important background of atmospheric gamma rays produced mainly by bremsstrahlung from secondary cascade electrons, but also by electron–positron annihilation, nuclear de-excitation, and $\pi^0$ decay. This background complicates the detection of a genuine geo-synchrotron signal and motivates a hybrid detection strategy incorporating the observation of other emissions (such as Cherenkov) for defining suitable triggering schemes. Another relevant process is synchrotron emission from cosmic-ray electron primaries. Detection of such emission was the primary goal of the CREST experiment~\cite{crest_experiment, crest_experiment2}, which sought to measure cosmic electrons in the \unit{TeV} domain to probe the diffusion coefficients of nearby accelerators, the imprint of local sources on the observed spectra, and potential contributions from dark-matter. Given that the CREST detector had geometric dimensions and an observational footprint comparable to those considered here, one would expect a substantially larger event rate from synchrotron emission by cosmic electrons than from the X-ray geo-synchrotron signal modeled in this work. Our simulation framework can be readily extended to model this scenario as well, and such studies will form part of our future work.

In summary, the framework developed here provides a fast method to estimate X-ray geo-synchrotron emission and its detectability from high-altitude platforms. With further refinement and expansion to include cosmic-electron synchrotron emission and improved modeling of young-shower electrons, we hope that our simulation framework can guide the design of future balloon and suborbital experiments targeting PeV-scale CR phenomena.

\bibliography{intro_references, references}

@book{aloisio_multiple_messengers,
    editor = "Aloisio, Roberto and Coccia, Eugenio and Vissani, Francesco",
    title = "{Multiple Messengers and Challenges in Astroparticle Physics}",
    doi = "10.1007/978-3-319-65425-6",
    isbn = "978-3-319-65423-2, 978-3-030-09739-4, 978-3-319-65425-6",
    publisher = "Springer",
    address = "Cham",
    year = "2018"
}

@article{snowmass_he_and_uhe_neutrinos,
    author = "Ackermann, Markus and others",
    title = "{High-energy and ultra-high-energy neutrinos: A Snowmass white paper}",
    eprint = "2203.08096",
    archivePrefix = "arXiv",
    primaryClass = "hep-ph",
    doi = "10.1016/j.jheap.2022.08.001",
    journal = "JHEAp",
    volume = "36",
    pages = "55--110",
    year = "2022"
}

@article{acceleration_and_propagation_of_uhecrs,
    author = "Aloisio, Roberto",
    title = "{Acceleration and propagation of ultra high energy cosmic rays}",
    eprint = "1707.08471",
    archivePrefix = "arXiv",
    primaryClass = "astro-ph.HE",
    doi = "10.1093/ptep/ptx115",
    journal = "PTEP",
    volume = "2017",
    number = "12",
    pages = "12A102",
    year = "2017"
}

@article{snowmass_companion_uhecr,
    author = "Coleman, A. and others",
    title = "{Ultra high energy cosmic rays The intersection of the Cosmic and Energy Frontiers}",
    eprint = "2205.05845",
    archivePrefix = "arXiv",
    primaryClass = "astro-ph.HE",
    reportNumber = "FERMILAB-PUB-22-413-PPD",
    doi = "10.1016/j.astropartphys.2023.102819",
    journal = "Astropart. Phys.",
    volume = "149",
    pages = "102819",
    year = "2023"
}

@inbook{cr_selected_topics,
    author = "Aloisio, Roberto and Blasi, Pasquale and De Mitri, Ivan and Petrera, Sergio",
    editor = "Aloisio, Roberto and Coccia, Eugenio and Vissani, Francesco",
    title = "{Selected Topics in Cosmic Ray Physics}",
    booktitle = "{Multiple Messengers and Challenges in Astroparticle Physics}",
    eprint = "1707.06147",
    archivePrefix = "arXiv",
    primaryClass = "astro-ph.HE",
    publisher="Springer International Publishing",
    doi = "10.1007/978-3-319-65425-6_1",
    pages = "1--95",
    year = "2018"
}

@article{lhaaso_all_particle,
    author = "Cao, Zhen and others",
    collaboration = "LHAASO",
    title = "{Measurements of All-Particle Energy Spectrum and Mean Logarithmic Mass of Cosmic Rays from 0.3 to 30~PeV with LHAASO-KM2A}",
    eprint = "2403.10010",
    archivePrefix = "arXiv",
    primaryClass = "astro-ph.HE",
    doi = "10.1103/PhysRevLett.132.131002",
    journal = "Phys. Rev. Lett.",
    volume = "132",
    number = "13",
    pages = "131002",
    year = "2024"
}

@article{icetop_all_particle,
    author = "Aartsen, M. G. and others",
    collaboration = "IceCube",
    title = "{Cosmic ray spectrum from 250 TeV to 10 PeV using IceTop}",
    eprint = "2006.05215",
    archivePrefix = "arXiv",
    primaryClass = "astro-ph.HE",
    doi = "10.1103/PhysRevD.102.122001",
    journal = "Phys. Rev. D",
    volume = "102",
    pages = "122001",
    year = "2020"
}

@article{grapes_all_particle,
    author = "Varsi, Fahim",
    title = "{Cosmic ray energy and composition measurements with GRAPES-3 and other experiments}",
    doi = "10.1140/epjs/s11734-025-01707-8",
    journal = "Eur. Phys. J. ST",
    volume = "234",
    number = "16",
    pages = "5021--5030",
    year = "2025"
}

@article{tale_all_particle,
    author = "Abbasi, R. U. and others",
    collaboration = "Telescope Array",
    title = "{The Cosmic-Ray Composition between 2 PeV and 2 EeV Observed with the TALE Detector in Monocular Mode}",
    eprint = "2012.10372",
    archivePrefix = "arXiv",
    primaryClass = "astro-ph.HE",
    doi = "10.3847/1538-4357/abdd30",
    journal = "Astrophys. J.",
    volume = "909",
    number = "2",
    pages = "178",
    year = "2021"
}

@article{auger_all_particle,
    author = "Halim, A. Abdul and others",
    collaboration = "Pierre Auger",
    title = "{Constraining models for the origin of ultra-high-energy cosmic rays with a novel combined analysis of arrival directions, spectrum, and composition data measured at the Pierre Auger Observatory}",
    eprint = "2305.16693",
    archivePrefix = "arXiv",
    primaryClass = "astro-ph.HE",
    reportNumber = "FERMILAB-PUB-24-0135-AD-CSAID-PPD-TD-V",
    doi = "10.1088/1475-7516/2024/01/022",
    journal = "JCAP",
    volume = "2024",
    number = "01",
    pages = "022",
    year = "2024"
}

@article{auger_highlights,
    author = "Salamida, Francesco",
    collaboration = "Pierre Auger",
    title = "{Highlights from the Pierre Auger Observatory}",
    eprint = "2312.14673",
    archivePrefix = "arXiv",
    primaryClass = "astro-ph.HE",
    reportNumber = "PoS(ICRC2023)016",
    doi = "10.22323/1.444.0016",
    journal = "PoS",
    volume = "ICRC2023",
    pages = "016",
    year = "2023"
}

@article{ta_all_particle,
    author = "Abu-Zayyad, T. and others",
    collaboration = "Telescope Array",
    title = "{The surface detector array of the Telescope Array experiment}",
    eprint = "1201.4964",
    archivePrefix = "arXiv",
    primaryClass = "astro-ph.IM",
    doi = "10.1016/j.nima.2012.05.079",
    journal = "Nucl. Instrum. Meth. A",
    volume = "689",
    pages = "87--97",
    year = "2013"
}

@article{shape_of_cr_spectrum,
    author = "Lipari, Paolo and Vernetto, Silvia",
    title = "{The shape of the cosmic ray proton spectrum}",
    eprint = "1911.01311",
    archivePrefix = "arXiv",
    primaryClass = "astro-ph.HE",
    doi = "10.1016/j.astropartphys.2020.102441",
    journal = "Astropart. Phys.",
    volume = "120",
    pages = "102441",
    year = "2020"
}

@article{euso_spb2,
    author = "Eser, Johannes and Olinto, Angela V. and Wiencke, Lawrence",
    collaboration = "JEM-EUSO",
    title = "{Overview and First Results of EUSO-SPB2}",
    eprint = "2308.15693",
    archivePrefix = "arXiv",
    primaryClass = "astro-ph.HE",
    doi = "10.22323/1.444.0397",
    journal = "PoS",
    volume = "ICRC2023",
    pages = "397",
    year = "2023"
}

@article{pueo,
    author = "Abarr, Q. and others",
    collaboration = "PUEO",
    title = "{The Payload for Ultrahigh Energy Observations (PUEO): a white paper}",
    eprint = "2010.02892",
    archivePrefix = "arXiv",
    primaryClass = "astro-ph.IM",
    doi = "10.1088/1748-0221/16/08/P08035",
    journal = "JINST",
    volume = "16",
    number = "08",
    pages = "P08035",
    year = "2021"
}

@article{pbr,
    author = "Heibges, Julia Burton",
    collaboration = "JEM-EUSO",
    title = "{POEMMA-Balloon with Radio: Mission Overview}",
    doi = "10.1088/1742-6596/3053/1/012012",
    journal = "J. Phys. Conf. Ser.",
    volume = "3053",
    number = "1",
    pages = "012012",
    year = "2025"
}

@ARTICLE{westfold_synchrotron_polarization,
       author = {{Westfold}, K.~C.},
        title = "{The Polarization of Synchrotron Radiation.}",
      journal = {\apj},
         year = 1959,
        month = {Jul},
       volume = {130},
        pages = {241},
          doi = {10.1086/146713},
}

@article{ap_synchrotron_theory_1,
    author = "Aloisio, Roberto and Blasi, Pasquale",
    title = "{Theory of synchrotron radiation. 1. Coherent emission from ensembles of particles}",
    eprint = "astro-ph/0201310",
    archivePrefix = "arXiv",
    doi = "10.1016/S0927-6505(02)00103-2",
    journal = "Astropart. Phys.",
    volume = "18",
    pages = "183--193",
    year = "2002"
}

@article{lafebre2009universality,
    author = "Lafebre, S. and Engel, R. and Falcke, H. and Horandel, J. and Huege, T. and Kuijpers, J. and Ulrich, R.",
    title = "{Universality of electron-positron distributions in extensive air showers}",
    eprint = "0902.0548",
    archivePrefix = "arXiv",
    primaryClass = "astro-ph.HE",
    doi = "10.1016/j.astropartphys.2009.02.002",
    journal = "Astropart. Phys.",
    volume = "31",
    pages = "243--254",
    year = "2009"
}

@article{nerling2006universality,
    author = "Nerling, Frank and Bluemer, J. and Engel, R. and Risse, M.",
    title = "{Universality of electron distributions in high-energy air showers: Description of Cherenkov light production}",
    eprint = "astro-ph/0506729",
    archivePrefix = "arXiv",
    doi = "10.1016/j.astropartphys.2005.09.002",
    journal = "Astropart. Phys.",
    volume = "24",
    pages = "421--437",
    year = "2006"
}

@techreport{heck1998corsika,
    author = "Heck, D. and Knapp, J. and Capdevielle, J. N. and Schatz, G. and Thouw, T.",
    title = "{CORSIKA: A Monte Carlo code to simulate extensive air showers}",
    reportNumber = "FZKA-6019",
    month = "2",
    year = "1998"
}

@book{rybicki2024radiative,
  title={Radiative processes in astrophysics},
  author={Rybicki, George B and Lightman, Alan P},
  year={2024},
  publisher={John Wiley \& Sons}
}

@article{hillas1982angular,
    author = "Hillas, A. M.",
    title = "{Angular and energy distributions of charged particles in electron-photon cascades in air}",
    doi = "10.1088/0305-4616/8/10/016",
    journal = "J. Phys. G",
    volume = "8",
    pages = "1461--1473",
    year = "1982"
}

@inproceedings{bergman2013efficient,
    author = "Bergman, Douglas",
    title = "{An Efficient Technique for the Reconstruction of Extensive Air Showers using Non-Imaging Cherenkov Measurements}",
    booktitle = "{33rd International Cosmic Ray Conference}",
    pages = "0983",
    year = "2013"
}

@article{kalos1954angular,
  title={The Angular Distribution of High Energy Electrons in Air Showers. I. Landau Approximation},
  author={Kalos, MH and Blatt, JM},
  journal={Australian Journal of Physics},
  volume={7},
  number={4},
  pages={543--569},
  year={1954},
  publisher={CSIRO Publishing}
}

@article{kamata1958lateral,
    author = "Kamata, Koichi and Nishimura, Jun",
    title = "{The Lateral and the Angular Structure Functions of Electron Showers}",
    doi = "10.1143/PTPS.6.93",
    journal = "Prog. Theor. Phys. Suppl.",
    volume = "6",
    pages = "93--155",
    year = "1958"
}

@article{greisenlateral,
    author = "Greisen, K.",
    title = "{Cosmic ray showers}",
    doi = "10.1146/annurev.ns.10.120160.000431",
    journal = "Ann. Rev. Nucl. Part. Sci.",
    volume = "10",
    pages = "63--108",
    year = "1960"
}

@article{cummings2021modeling1,
    author = "Cummings, Austin and Aloisio, Roberto and Eser, Johannes and Krizmanic, John",
    title = "{Modeling the optical Cherenkov signals by cosmic ray extensive air showers directly observed from suborbital and orbital altitudes}",
    eprint = "2105.03255",
    archivePrefix = "arXiv",
    primaryClass = "astro-ph.IM",
    doi = "10.1103/PhysRevD.104.063029",
    journal = "Phys. Rev. D",
    volume = "104",
    number = "6",
    pages = "063029",
    year = "2021"
}

@article{cummings2021modeling2,
    author = "Cummings, A. L. and Aloisio, R. and Krizmanic, J. F.",
    title = "{Modeling of the Tau and Muon Neutrino-induced Optical Cherenkov Signals from Upward-moving Extensive Air Showers}",
    eprint = "2011.09869",
    archivePrefix = "arXiv",
    primaryClass = "astro-ph.HE",
    doi = "10.1103/PhysRevD.103.043017",
    journal = "Phys. Rev. D",
    volume = "103",
    number = "4",
    pages = "043017",
    year = "2021"
}

@book{atmosphere1976us,
  title={US standard atmosphere},
  author={Atmosphere, US Standard},
  year={1976},
  publisher={National Oceanic and Atmospheric Administration}
}

@misc{nist,
  author = {Martin Berger and J Hubbell and Stephen Seltzer and J Coursey and D Zucker},
  title = {XCOM: Photon Cross Section Database (version 1.2)},
  year = {1999},
  month = {1999-01-01},
  publisher = {http://physics.nist.gov/xcom}
}

@article{raspass_simulation,
    author = "Tueros, Matias Jorge",
    title = "{Simulation of radiopulses from Atmosphere-Skimming Extensive Air Showers with ZHAireS-RASPASS}",
    doi = "10.22323/1.424.0056",
    journal = "PoS",
    volume = "ARENA2022",
    pages = "056",
    year = "2023"
}

@article{raspass_theory,
    author = "Tueros, Mat{\'\i}as and Cabana-Freire, Sergio and {\'A}lvarez-Mu{\~n}iz, Jaime",
    title = "{Radio emission from atmosphere-skimming cosmic ray showers in high-altitude balloon-borne experiments}",
    eprint = "2409.13141",
    archivePrefix = "arXiv",
    primaryClass = "astro-ph.IM",
    doi = "10.1088/1475-7516/2025/01/112",
    journal = "JCAP",
    volume = "01",
    pages = "112",
    year = "2025"
}

@article{giller2004energy,
    author = "Giller, M. and Wieczorek, G. and Kacperczyk, A. and Stojek, H. and Tkaczyk, W.",
    title = "{Energy spectra of electrons in the extensive air showers of ultra-high energy}",
    doi = "10.1088/0954-3899/30/2/009",
    journal = "J. Phys. G",
    volume = "30",
    pages = "97--105",
    year = "2004"
}

@article{lipari_universality,
    author = "Lipari, Paolo",
    title = "{The Concepts of 'Age' and 'Universality' in Cosmic Ray Showers}",
    eprint = "0809.0190",
    archivePrefix = "arXiv",
    primaryClass = "astro-ph",
    doi = "10.1103/PhysRevD.79.063001",
    journal = "Phys. Rev. D",
    volume = "79",
    pages = "063001",
    year = "2009"
}

@article{Guepin:2021ljb,
    author = "Gu{\'e}pin, Claire and Aloisio, Roberto and Anchordoqui, Luis A. and Cummings, Austin and Krizmanic, John F. and Olinto, Angela V. and Reno, Mary Hall and Venters, Tonia M.",
    title = "{Indirect dark matter searches at ultrahigh energy neutrino detectors}",
    eprint = "2106.04446",
    archivePrefix = "arXiv",
    primaryClass = "hep-ph",
    doi = "10.1103/PhysRevD.104.083002",
    journal = "Phys. Rev. D",
    volume = "104",
    number = "8",
    pages = "083002",
    year = "2021"
}

@article{Aloisio:2007bh,
    author = "Aloisio, R. and Tortorici, F.",
    title = "{Super Heavy Dark Matter and UHECR Anisotropy at Low Energy}",
    eprint = "0706.3196",
    archivePrefix = "arXiv",
    primaryClass = "astro-ph",
    doi = "10.1016/j.astropartphys.2008.02.005",
    journal = "Astropart. Phys.",
    volume = "29",
    pages = "307--316",
    year = "2008"
}

@article{Aloisio:2025nts,
  title={Constraining Super-Heavy Dark Matter with the KM3-230213A Neutrino Event},
  author={Aloisio, Roberto and Ambrosone, Antonio and Evoli, Carmelo},
  journal={arXiv preprint arXiv:2508.08779},
  year={2025}
}

@article{PierreAuger:2023yym,
    author = "Abdul Halim, Adila and others",
    collaboration = "Pierre Auger, Telescope Array",
    title = "{Depth of maximum of air-shower profiles: testing the compatibility of the measurements at the Pierre Auger Observatory and the Telescope Array}",
    doi = "10.22323/1.444.0249",
    journal = "PoS",
    volume = "ICRC2023",
    pages = "249",
    year = "2023"
}

@article{PierreAuger:2025xxv,
    author = "Bergman, Douglas R. and others",
    collaboration = "Pierre Auger, Telescope Array",
    title = "{Updated comparison of the UHECR energy spectra measured by the Pierre Auger Observatory and the Telescope Array}",
    eprint = "2509.05530",
    archivePrefix = "arXiv",
    primaryClass = "astro-ph.HE",
    reportNumber = "PoS ( ICRC2025 ) 480, PoS ( ICRC2025 ) 381",
    doi = "10.22323/1.501.0381",
    journal = "PoS",
    volume = "ICRC2025",
    pages = "381",
    year = "2025"
}

@article{Colin:2006nj,
    author = "Colin, Pierre and Chukanov, A. and Grebenyuk, V. and Naumov, D. and Nedelec, P. and Nefedov, Y. and Onofre, A. and Porokhovoi, S. and Sabirov, B. and Tkatchev, L.",
    title = "{Measurement of air and nitrogen fluorescence light yields induced by electron beam for UHECR experiments}",
    eprint = "astro-ph/0612110",
    archivePrefix = "arXiv",
    doi = "10.1016/j.astropartphys.2006.11.008",
    journal = "Astropart. Phys.",
    volume = "27",
    pages = "317--325",
    year = "2007"
}

@article{sullivan1971geometric,
  title={Geometric factor and directional response of single and multi-element particle telescopes},
  author={Sullivan, JD},
  journal={Nuclear Instruments and methods},
  volume={95},
  number={1},
  pages={5--11},
  year={1971},
  publisher={Elsevier}
}

@article{CORSIKA:2025zlj,
    author = "Alameddine, Jean-Marco and others",
    collaboration = "CORSIKA",
    title = "{Overview of the CORSIKA 8 astroparticle simulation framework}",
    doi = "10.22323/1.501.0371",
    journal = "PoS",
    volume = "ICRC2025",
    pages = "371",
    year = "2025"
}

@article{crannell_MC,
    author = "Crannell, C. J. and Ormes, J. F.",
    title = "{Geometrical-factor determination using a monte carlo approach}",
    doi = "10.1016/0029-554X(71)90357-0",
    journal = "Nucl. Instrum. Meth.",
    volume = "94",
    pages = "179--183",
    year = "1971"
}

@article{ling_background,
author = {Ling, James C.},
title = {A semiempirical model for atmospheric $\gamma$ rays from 0.3 to 10 MeV at $\lambda$ = 40°},
journal = {Journal of Geophysical Research (1896-1977)},
volume = {80},
number = {22},
pages = {3241-3252},
doi = {https://doi.org/10.1029/JA080i022p03241},
eprint = {https://agupubs.onlinelibrary.wiley.com/doi/pdf/10.1029/JA080i022p03241},
year = {1975}
}

@inproceedings{crest_experiment,
    author = "Schubnell, Michael and others",
    title = "{The Cosmic Ray Electron Synchrotron Telescope (CREST) Experiment}",
    booktitle = "{30th International Cosmic Ray Conference}",
    volume = "2",
    pages = "305--308",
    month = "7",
    year = "2007"
}

@article{crest_experiment2,
    author = "Coutu, S. and others",
    editor = "Govoni, Pietro and Marrocchesi, Pier Simone and Navarria, Francesco-Luigi and Paganoni, Marco and Perrotta, Andrea",
    title = "{Searching for TeV cosmic electrons with the CREST experiment}",
    doi = "10.1016/j.nuclphysbps.2011.04.022",
    journal = "Nucl. Phys. B Proc. Suppl.",
    volume = "215",
    pages = "250--254",
    year = "2011"
}

@article{krizmanic_initial,
    author = "Krizmanic, John F. and Cummings, Austin L. and Garcia, Fred A. B. and Mitchell, John W.",
    title = "{Extensive Air Shower (EAS) Development in the Upper Atmosphere: a unique environment to measure the EAS properties}",
    doi = "10.22323/1.444.0524",
    journal = "PoS",
    volume = "ICRC2023",
    pages = "524",
    year = "2023"
}

@article{IceCube_2019hmk,
    author = "Aartsen, M. G. and others",
    collaboration = "IceCube",
    title = "{Cosmic ray spectrum and composition from PeV to EeV using 3 years of data from IceTop and IceCube}",
    eprint = "1906.04317",
    archivePrefix = "arXiv",
    primaryClass = "astro-ph.HE",
    doi = "10.1103/PhysRevD.100.082002",
    journal = "Phys. Rev. D",
    volume = "100",
    number = "8",
    pages = "082002",
    year = "2019"
}

@article{Apel_2013uni,
    author = "Apel, W. D. and others",
    title = "{KASCADE-Grande measurements of energy spectra for elemental groups of cosmic rays}",
    eprint = "1306.6283",
    archivePrefix = "arXiv",
    primaryClass = "astro-ph.HE",
    doi = "10.1016/j.astropartphys.2013.06.004",
    journal = "Astropart. Phys.",
    volume = "47",
    pages = "54--66",
    year = "2013"
}

@article{Alvarez-Muniz_2018owm,
    author = "Alvarez-Mu{\~n}iz, Jaime and Carvalho, Washington R. and Cummings, Austin L. and Payet, K{\'e}vin and Romero-Wolf, Andr{\'e}s and Schoorlemmer, Harm and Zas, Enrique",
    title = "{Comprehensive approach to tau-lepton production by high-energy tau neutrinos propagating through the Earth}",
    eprint = "1707.00334",
    archivePrefix = "arXiv",
    primaryClass = "astro-ph.HE",
    doi = "10.1103/PhysRevD.97.023021",
    journal = "Phys. Rev. D",
    volume = "97",
    number = "2",
    pages = "023021",
    year = "2018",
    note = "[Erratum: Phys.Rev.D 99, 069902 (2019)]"
}
\end{document}